\documentclass[pdflatex,sn-mathphys-num]{sn-jnl}% Math and Physical Sciences Numbered Reference Style
\usepackage{graphicx}%
\usepackage{multirow}%
\usepackage{amsmath,amssymb,amsfonts}%
\usepackage{amsthm}%
\usepackage{mathrsfs}%
\usepackage[title]{appendix}%
\usepackage{xcolor}%
\usepackage{textcomp}%
\usepackage{manyfoot}%
\usepackage{booktabs}%
\usepackage{algorithm}%
\usepackage{algorithmicx}%
\usepackage{algpseudocode}%
\usepackage{listings}%
\theoremstyle{thmstyleone}%
\theoremstyle{thmstyletwo}%

\theoremstyle{thmstylethree}%

\def\prl{Physical Review Letters}
\def\apjl{Astrophysical Journal Letters}
\def\apj{Astrophysical Journal}

\def\ssr{Space Science Reviews}

\def\grl{Geophysical Research Letters}

\def\mnras{Monthly Notices of the Royal Astronomical Society}

\def\aap{Astronomy and Astrophysics}

\begin{document}

\title[Article Title]{Particle Injection Problem in Magnetic Reconnection and Turbulence}

%%=============================================================%%
%% GivenName	-> \fnm{Joergen W.}
%% Particle	-> \spfx{van der} -> surname prefix
%% FamilyName	-> \sur{Ploeg}
%% Suffix	-> \sfx{IV}
%% \author*[1,2]{\fnm{Joergen W.} \spfx{van der} \sur{Ploeg} 
%%  \sfx{IV}}\email{iauthor@gmail.com}
%%=============================================================%%

\author*[1]{\fnm{Fan} \sur{Guo}}\email{guofan@lanl.gov} 

\author[1,2]{\fnm{Omar} \sur{French}}\email{omar.french@colorado.edu}
\author[3]{\fnm{Qile} \sur{Zhang}}

\author[1]{\fnm{Xiaocan} \sur{Li}}\email{xiaocanli@lanl.gov}

\author[1]{\fnm{Jeongbhin} \sur{Seo}}\email{jseo@lanl.gov}

%\equalcont{These authors contributed equally to this work.}

\affil[1]{Los Alamos National Laboratory, Los Alamos, New Mexico 87545, USA}

\affil[2]{Center for Integrated Plasma Studies, Department of Physics, 390 UCB, University of Colorado, Boulder, CO 80309, USA}

\affil[3]{University of Maryland, College Park}

%%==================================%%
%% Sample for unstructured abstract %%
%%==================================%%

\abstract{Magnetic reconnection and turbulence in magnetically-dominated environments have been proposed as important nonthermal particle acceleration mechanisms that generate high energy particles and associated emissions. While the acceleration to high energy that produces the power-law energy distribution has drawn strong interest, recent studies actively discuss pre-acceleration, or injection, to a sufficient energy for a sustained and prolonged Fermi-like acceleration. The injection process is important for determining the fraction of nonthermal particles and energy partition between thermal and nonthermal particles. We review recent advances in understanding the injection mechanisms responsible for populating these nonthermal power-law spectra, and conclude with an outlook for studies and applications of injection models.}


%%\pacs[JEL Classification]{D8, H51}

%%\pacs[MSC Classification]{35A01, 65L10, 65L12, 65L20, 65L70}

\maketitle

\section{Introduction} \label{sec:introduction}

The origin of energetic charged particles and associated emissions in the Universe is a long-standing problem in high-energy astrophysics and plasma astrophysics. Examples include those in supernova remnants, solar flares, heliospheric shocks, Earth's magnetotail, pulsar wind nebulae, and jets and accretion disks surrounding black holes \citep{Blandford_1987,Abeysekara2017,Chen2020,Reames1999,Li2021,Oka2023,Guo2024,Oka2025,Drake2025}. The high energy particle distribution often takes the form of a power law that contains important information on particle acceleration. Therefore, to explain high-energy particles and emissions, it is essential to understand how power-law distributions of nonthermal particles are initiated and further extended in high energy. The former process is mainly concerned with how particles are accelerated to the lower energy bound of the power-law distribution (``injection''), while the latter mostly studies the nonthermal particle acceleration to higher energies, leading to the power-law distribution itself. The acceleration to high energy is often thought as a Fermi-like process \citep{Fermi1949,Bell_1978,Blandford_1987,Drury1999,Guo2014,Comisso2018,Lemoine2019,Lemoine_2020}, but the acceleration mechanisms involved in the injection process can often be more complicated \citep{French_2023,Guo2013,Zhang2024ApJ,Comisso2019,Caprioli2015,Sironi2022,Guo2024}. This review paper focuses on the injection process recently studied in magnetic reconnection and turbulence.

The injection problem can be considered from two aspects:
(1) Under what conditions are particles eligible to participate in a continual, power-law forming acceleration process? 
(2) What physical mechanisms are responsible for granting particles this eligibility? 
Regarding the first aspect, \citet{Fermi1949} first formulated an injection kinetic energy threshold~$\varepsilon_{\rm inj}$ that a particle must surpass, so that particles with kinetic energy~$\varepsilon > \varepsilon_{\rm inj}$ can experience Fermi acceleration and are accelerated into a power-law energy distribution. 
Physically, this may happen as the scales of particle gyromotion (which are proportional to the particle energy) become large enough to facilitate stochastic acceleration off plasma structures~\citep{Lemoine_2020}. Thus the common solution to the first question is to determine the injection energy~$\varepsilon_{\rm inj}$ and relate it to the key properties of the process. Accordingly, the second question is to identify, understand, and evaluate the importance of the various ``injection mechanisms", i.e., mechanisms that accelerate particles from a thermal bath to the injection energy~$\varepsilon_{\rm inj}$ that initiate the nonthermal power-law spectrum. % In addition, under what conditions are these injection mechanisms important.

Since the injection process describes how particles are accelerated into the nonthermal range, it determines the fraction of nonthermal particles and the energy contained in the nonthermal distribution, i.e., the efficiency of nonthermal acceleration \citep{French_2023,Hoshino_2023}. The injection process leads to energy partition between thermal and nonthermal components, and the injection process of different species determines their relative abundance in the nonthermal distribution. This is important for studying the acceleration of different species \citep{Guo2016,Zhang2021,Comisso2022,Zhang2024a}. 

\begin{figure*}[htp!]
    \centering
    \includegraphics[width = \textwidth]{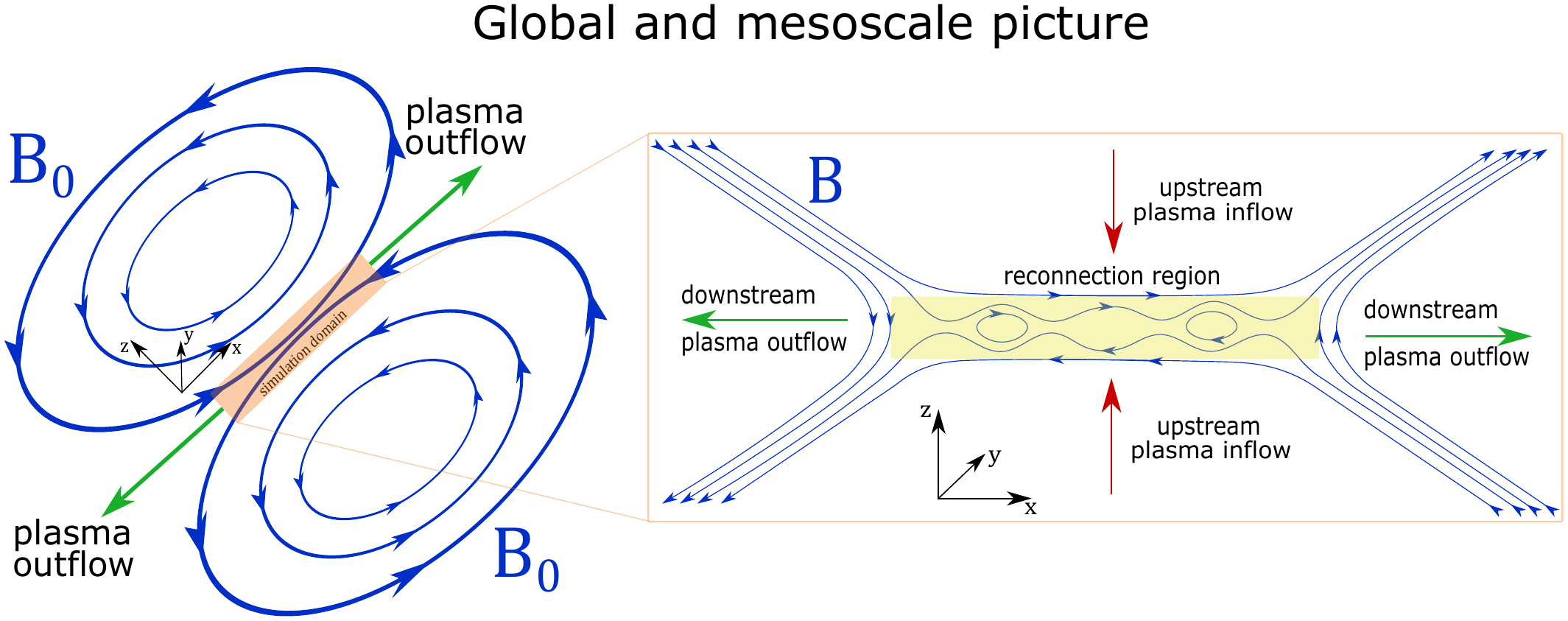}
    \caption{Cartoons of global and mesoscale reconnection configurations of magnetic reconnection, adapted from~\citet{French_2023} [reproduced by permission of the AAS].}
    \label{fig:global_mesoscale_cartoon}
\end{figure*}

Despite the importance of the ``injection problem'' to nonthermal particle acceleration, its solution has not been fully understood. The injection problem has been extensively discussed in collisionless shocks \citep{Ellison1990,Giacalone1992,kirk2001,Guo2013,Caprioli2015,Parsons_2023}, but has not been comprehensively studied in the context of other plasma processes. Recently, the formation of power-law distributions and the injection problem has been discussed in magnetic reconnection in the magnetically dominated regime ($\sigma_h \equiv B_0^2/4\pi h \gg 1$, where~$B_0$ is the ambient upstream magnetic field and~$h$ is the plasma enthalpy) \citep{Zenitani2001,Zenitani2005,Cerutti2013,Sironi2014,Melzani2014,Guo2014,Guo2015,Guo2019,Werner2016,Werner2017,Werner2018,Schoeffler2019,Mehlhaff2020,Hakobyan2021,Sironi2022,Li2023,French_2023,HaoZhang_2023}. This has been recognized to be important for jets from active galactic nuclei (AGN) \citep{Romanova1992, Giannios2009,Giannios2010,Sironi2015,Blandford2019,Zhang2020} and pulsar wind nebulae (PWNe) [see \citet[][for focused reviews]{Hoshino2012,Kagan2015,Guo2020,Guo2024}].
%These questions of \textit{particle injection}---despite predicating the universal Fermi process that generates nonthermal power-law spectra and, subsequently, high-energy emissions---have been unresolved in a variety of astrophysical contexts, including those which are magnetically-dominated  such as 
%\citep{Rees1974,Uzdensky2011b,Sironi2011,Komissarov2011,Cerutti2013,Sironi2017,Lyutikov2018,Cerutti2020,Lu2021}. 
These studies have also triggered more reconnection studies in transrelativistic ($\sigma_h \sim 1$) \citep{Ball2018,Ball2019,Kilian2020,Werner2021,Li2023} and nonrelativistic ($\sigma_h \ll 1$ but plasma $\beta \ll 1$)  regimes \citep{Drake2006,Oka_2010,Dahlin2014,Dahlin2016,Dahlin2017,Li_2015,Li2017,Li2019a,Li2019b}, such as the solar corona and black hole accretion disk coronae~\citep{Chen2020,Hoshino_2015,Ripperda_2020,Nathanail_2022,Lin_2023}. In addition, magnetized turbulence has been suggested to strongly accelerate particles \citep{Chandran_2000,Zhdankin_2017,Zhdankin_2018,Zhdankin_2019,Comisso2018,Comisso2019,Wong2020}. 
The Fermi process has been considered to be a very generic mechanism \citep{Lemoine_2020} that commonly exists in magnetic reconnection and turbulence \citep{Drake2006,Guo2014,Comisso2019}. Therefore it is important to achieve a comprehensive understanding of the injection process in different processes.

In the context of relativistic magnetic reconnection, how exactly the nonthermal particles are generated is still a subject of debate. Studies have shown that the acceleration to high energy has to be via the large-scale $\textbf{u}\times \textbf{B}$ electric field, and the mechanism of power-law formation does not require the non-ideal electric field \citep{Guo2014,Guo2019,Hakobyan2021}. However, the low-energy injection is still unclear. \citet{Sironi2014} and \citet{Sironi2022} claim that the acceleration in diffusion regions with the electric field larger than the magnetic field $E>B$ is essential (following \citet{Zenitani2001}). They argue that most of the nonthermal particles went through the $E>B$ regions and gain significant amount of energy.  However, this acceleration has been demonstrated to be inefficient and cannot achieve injection for most of particles \citep{Guo2019,Guo2023_comment}. \citet{French_2023} studied three injection mechanisms (direct acceleration, Fermi reflection, and pickup process) and quantified their contributions for a variety of different guide-field strengths and system sizes. Further discussion has included the non-ideal electric field or parallel electric field that extends further beyond the diffusion region \citep{Guo2019,French_2023,Totorica_2023}.  Using a spectral fitting procedure, the injection energy~$\gamma_{\rm inj}$ has been measured directly~\citep{French_2023,Singh_2024,French_2025}, as well as its empirical dependence on guide-field strength~$b_g \equiv B_g/B_0$ (where~$B_g$ is the out-of-plane guide magnetic field)~\citep{French_2023}, upstream magnetization~\citep{French_2025}, and system size~\citep{French_2023,Singh_2024}. Recent studies have shown that the work done to a particle by a single Fermi reflection off a reconnection exhaust aligns closely with~$\varepsilon_{\rm inj}$, suggesting that Fermi reflection may explain these measurements \citep{Qile_2024,French_2025}. 
\citet{Li2023} demonstrated that particle energy spectra from relativistic reconnection can be well quantified by injection, acceleration, and escape processes. %In Section~\ref{sec:rel_recon}, we discuss the injection problem in relativistic magnetic reconnection. 

In addition, concurrent discussion on injection problems in magnetic reconnection in the transrelativistic and nonrelativistic regimes \citep{Ball2019,Kilian2020,Zhang2021} and turbulence in the relativistic regime \citep{Comisso2019,Singh_2024} have initiated very complementary studies. %We discuss them in Sections~\ref{sec:nonrel_recon} and~\ref{sec:turb}.
Assembling these studies, researchers have started to make significant progress toward a more comprehensive and unified understanding of nonthermal particle acceleration and the roles of the injection problem. 

A major goal of the particle acceleration studies is to understand and predict the feature of nonthermal distribution in astrophysical systems. Fig. \ref{fig:global_mesoscale_cartoon} visualizes the relation between the reconnection layer and the larger global scale. The reconnection region dissipates magnetic energy and accelerates particles to high energy, and feed energetic particles to other regions in the global domain. Predicting the nonthermal distribution in the macrosale reconnection region and global domain is challenging because of the prohibiting scale separation between the system size and the kinetic scales. It is therefore impossible for conventional kinetic simulations methods to model the global problem. Recently, a series of studies have attempted to model particle acceleration and transport in large-scale magnetohydrodynamic reconnection simulations \citep{Li2018b,Arnold2021,Seo2024,Pino2005,Murtas2024,Yin2024}. A critical input of these models is an injection model that determines where, when, and how many particles are injected. It is therefore important to evaluate the injection process as a step toward achieving macroscopic particle acceleration models. %\textcolor{red}{Due to the great separation between kinetic and global scales present in astrophysical systems and the great computational cost of particle-in-cell (PIC) simulations, it is infeasible to self-consistently evolve astrophysical plasmas with PIC simulations. Therefore, relatively inexpensive fluid simulations are often employed to model astrophysical systems, in which NTPA is accounted by embedding \textit{injection models} onto regions where NTPA is expected to occur, i.e. extended current sheets that are likely reconnection sites. We illustrate current sheets in a global and mesoscale context of a reconnecting current sheet in Figure~\ref{fig:global_mesoscale_cartoon}. }

In the past two decades or so, kinetic particle-in-cell (PIC) simulations have greatly advanced our understanding of particle acceleration in magnetic reconnection and turbulence by providing a robust first principles description of plasma dynamics and particle acceleration. Recent PIC analyses that examine a large number of tracer particles have started to address particle acceleration issues directly and collectively. 
In this paper, we review recent progress on particle injection in the context of magnetic reconnection in the relativistic regime \citep{Sironi2022,Guo2022_comment,Totorica_2023,French_2023,French_2025} (Section~\ref{sec:rel_recon}), magnetic reconnection in the transrelativistic and nonrelativistic \citep{Ball2019,Kilian2020,Qile_2024} regime (Section~\ref{sec:nonrel_recon}), and relativistic magnetic turbulence \citep{Comisso2019,Singh_2024} (Section~\ref{sec:turb}). We discuss our outlook for injection studies and astrophysical implications in Section~\ref{sec:discussion}. 

% Several nonthermal particle acceleration mechanisms have been studied theoretically in the context of magnetic reconnection, such as the ``direct" acceleration by the parallel electric field with a finite guide magnetic field (i.e., a finite non-reversing, out-of-plane component of the magnetic field) near X-points \citep{Larrabee2003,Zenitani2005,Zenitani2008,Cerutti2013,Cerutti2014,Ball2019}, Speiser orbits in the case of zero guide field \citep{Speiser1965,Hoshino2001,Zenitani2001,Uzdensky2011,Cerutti2012,Cerutti2013,Cerutti2014,Nalewajko2015,Uzdensky2022}, Fermi acceleration \citep{Fermi1949,Drake2006,Giannios2010,Guo2014,Guo2015,Dahlin2014,Qile_2021}, parallel electric field acceleration in the exhaust region \citep{Egedal2013,Zhang2019}, and acceleration by the pickup process in the exhaust region \citep{Drake2009,Sironi2020,French_2023,Chernoglazov2023}.

\section{Injection Problem in Relativistic magnetic reconnection} \label{sec:rel_recon}

Over the past decade, significant strides have been made in understanding particle acceleration in the relativistic regime of collisionless magnetic reconnection.  A group of seminal works  have demonstrated that relativistic magnetic reconnection is efficient at producing hard nonthermal power-law spectra~$\gamma^{-p}$ with~$p$ as small as $p \sim 1$ \citep{Zenitani2001,Sironi2014,Guo2014,Werner2016}.  These results have elevated the candidacy of relativistic reconnection as a progenitor for high-energy emissions in a variety of astrophysical environments.  The acceleration in the high $\sigma$ limit can be very efficient, with more than $10\%$ particles and $50\%$ energy residing in the nonthermal range \citep{French_2023}, suggesting a very efficient mechanism for pumping thermal particles to nonthermals.%, including jets from active galactic nuclei \citep{Giannios2009,Sironi2015,Blandford2019,Zhang2020} and pulsar wind nebulae (PWNe) \citep{Uzdensky2011b,Sironi2011,Komissarov2011,Cerutti2013,Sironi2017,Lyutikov2018,Cerutti2020,Lu2021}. 

While the nonthermal power-law spectra from relativistic reconnection have been well recognized, how they form remains a topic of debate. The generation of  power-law energy distributions favors acceleration mechanisms that scale with particle energy \citep{Guo2014,Guo2024}. In magnetic reconnection, Fermi-like acceleration within the contracting and merging magnetic islands can provide such a mechanism \citep{Drake2006,Guo2014,Guo2015,Guo2019}. On the other hand, any Fermi process may require particles already energetic enough to participate in this sustained acceleration period, as we mentioned above. This section focuses on discussing recent studies on low energy injection during relativistic magnetic reconnection.

\subsection{Are Diffusion Regions ($E>B$ regions) Important for Accelerating or Injecting Particles?} 

\begin{figure*}[htp!]
    \centering
    \includegraphics[width = \textwidth]{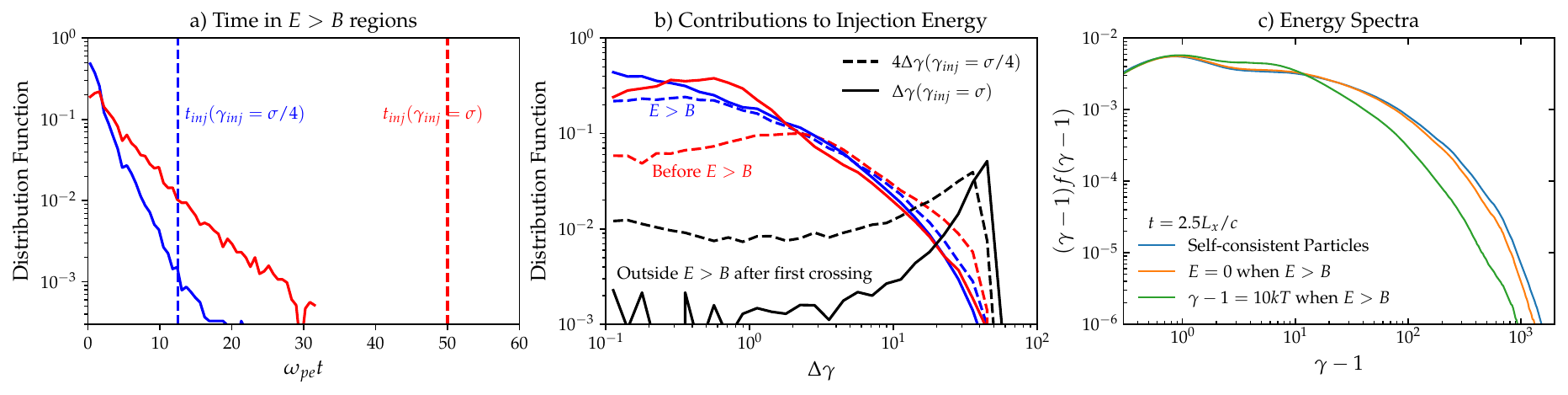}
    \caption{a) Distribution function for the time duration of $E > B$ particles within $E > B$ regions. The estimated minimum time for achieving injection are labeled assuming the reconnection rate is $R =0.1$; b) Distribution of energy gain during injection (for $E > B$ particles) for $E > B$ regions (blue), before $E > B$ crossing (red), and outside $E > B$ regions after the first crossing (black); c) Spectra for self-consistent particles (blue), test-particles that do not see electric fields in $E > B$ regions (red) and test-particles with a resetting energy approach (green; resembling Sironi \citep{Sironi2022}).}
    \label{fig:EgtB}
\end{figure*}

Whether the $E>B$ regions play a dominant role during the formation of the power-law distribution in a large-scale reconnection layer is controversial.
\citet{Sironi2014} suggest that particle acceleration within~$E>B$ regions is essential for producing the power-law spectrum and that the associated spectral index does not ``appreciably change'' during the further acceleration. In contrast, \citet{Guo2014,Guo2015} proposed that the power-law distribution is produced by a Fermi-like process and that the injection is sourced from the reconnection inflow. By tracing a large number of particles and the electromagnetic fields they ``see'' at every numerical time step in PIC simulations, \citet{Guo2019} have examined the acceleration in the $E>B$ regions and found the contribution of $E>B$ to the overall acceleration is very small compared to the total energy gain. In addition, the power-law indices between the diffusion region and outside the diffusion region appreciably changed owed to further Fermi-like acceleration in the broader reconnection layer. % is responsible for the change.  

It is also unclear whether $E>B$ regions are the dominant mechanism for injecting particles to the lower bound of the power law spectra during relativistic reconnection.
\citet{Sironi2022} argues that $E>B$ regions host and accelerate most particles that achieve injection. They claim that nearly all injected particles have to go through the $E>B$ regions. They also show if test particles are reset to low energies whenever they cross $E>B$ regions, injection is suppressed. These claims have been reexamined by \citet{Guo2023_comment}, who found a significant fraction of particles are injected without needing $E>B$ crossings. The discrepancy between the two correlation studies is that \citet{Sironi2022}'s analysis labels all particles that ever crossed $E>B$ regions during the entire simulation. In other words, the \citet{Sironi2022} study also includes particles going through the $E>B$ regions after injection. Even for the injected particles with $E>B$ crossings, \citet{Guo2023_comment} find that $E>B$ regions contribute very little to injection. Here we show results from a typical pair-plasma PIC simulations with no guide field~$b_g \sim 0$ and~$\sigma_0 \equiv B_0^2/4\pi n_0 m_e c^2 \sim 50$. Fig.~\ref{fig:EgtB}a) shows the distribution of time particles spend in $E>B$ regions before injection, excluding particles that did not encounter~$E>B$. The time can be used to infer the limit of acceleration by the reconnection electric field in the $E>B$ regions 
\begin{equation} 
\Delta \gamma_{E>B} \lesssim \int q r B_0 c dt / (m_ec^2),
\end{equation}
where the reconnection rate~$r \sim 0.1$. For~$\sigma = 50$, a duration of~$\omega_{\rm pe}t_{\rm inj}\gtrsim 50$ is needed for~$\gamma_{\rm inj}=\sigma$ ($\omega_{\rm pe}t_{\rm inj}\gtrsim 12.5$ if~$\gamma_{\rm inj}=\sigma/4$). The blue and red vertical dashed lines show the estimated times needed for particles to achieve the injection assuming for two different injection energy $\gamma_{\rm inj} = \sigma/4$ and $\gamma_{\rm inj} = \sigma$, respectively. Since $E>B$ regions host particles for a duration too short to inject most particles, they cannot be the primary injection mechanism. 

Fig.~\ref{fig:EgtB}b) shows the distribution of particle energy gain (during injection) in the $E>B$ regions (blue), before $E>B$ crossings (red), and outside $E>B$ regions after the first $E>B$ crossing (black). This is done by tracing particle trajectories at every time step during the simulation. Consistent with Fig.~\ref{fig:EgtB}a), the acceleration in $E>B$ regions is too little for direct injections of most particles. Interestingly, energization before any $E>B$ regions is comparable to $E>B$ acceleration. This suggests that $E>B$ acceleration is not unique in pre-accelerating particles before prolonged acceleration. 
Fig.~\ref{fig:EgtB}b) also shows that most acceleration during injection occurs outside $E>B$ regions. In Fig.~\ref{fig:EgtB}c), we evolve a test-particle component in the simulation that does not “see” the electric field in $E>B$ regions (so no acceleration during each crossing), and find $84\%$ ($94\%$) for $\gamma_{\rm inj}=\sigma$($\sigma/4$) compare to self-consistent particles are still injected. There is no significant difference between energy spectra of the test-particles and self-consistent particles. In contrast, when particle energies are reset to a low energy of during $E>B$ crossings (resembling \citet{Sironi2022}), particle injection is suppressed. This difference is because \citet{Sironi2022}'s resetting energy approach removes the acceleration before and between $E>B$ crossings as particles can cross $E>B$ regions multiple times.
\citet{Guo2023_comment} showed that \citet{Sironi2022} falsely quantified the number of particles going through the diffusion region and acceleration within the diffusion region.

\subsection{Injection Mechanisms} 

\begin{figure*}[htp!]
    \centering
    \includegraphics[width = \textwidth]{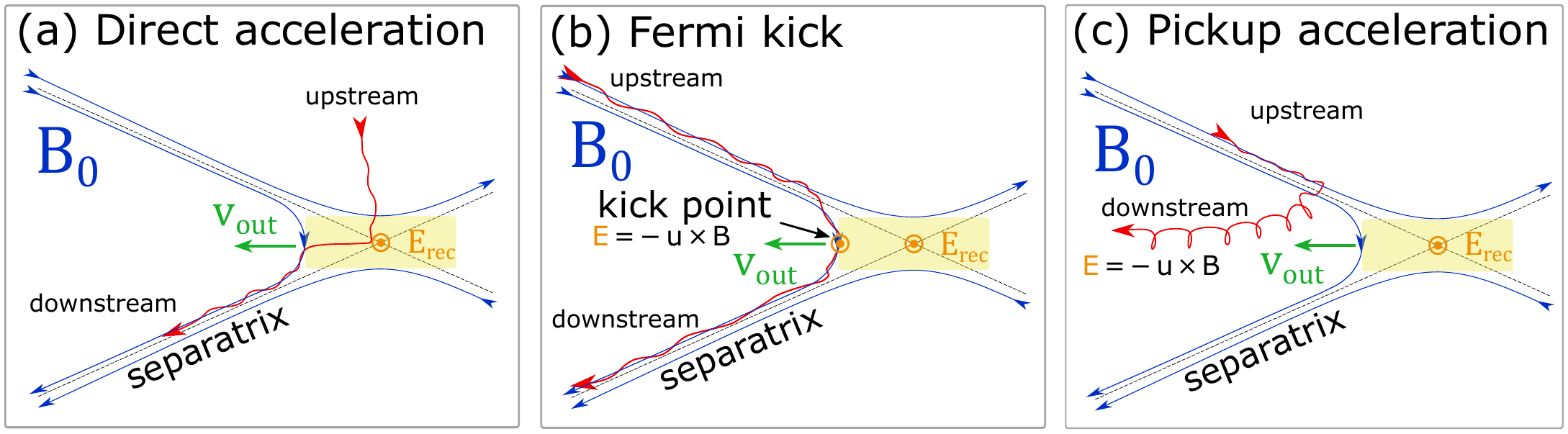}
    \caption{Sketches of several particle injection mechanisms surrounding a reconnection region, adapted from~\citet{French_2023}. In each panel, $B_0$ is the reconnecting magnetic field, $E_{\rm rec}$ is the reconnection electric field, and~$v_{\rm out} \simeq v_{\rm Ax}$ is the reconnection outflow speed, approximately equal to the in-plane Alfv\'en speed. (a) Injection by direct acceleration from the reconnection electric field surrounding an X-point. (b) Injection by a Fermi ``kick" downstream of the X-point. (c) Injection by the pickup process.}
    \label{fig:mechanism_cartoon}
\end{figure*}

The injection process can be attributed to several injection mechanisms. 
French et al. \cite{French_2023,French_2025} studied particle injection with different guide fields and upstream magnetizations during relativistic reconnection, considering three different acceleration mechanisms (see Figure~\ref{fig:mechanism_cartoon}):

\begin{enumerate}[leftmargin=*]
    \item Direct acceleration by the non-ideal electric field that surrounds reconnection X-points \citep{Larrabee2003,Zenitani2005,Zenitani2008,Cerutti2013,Cerutti2014,Ball2019,Sironi2022,Totorica_2023}, illustrated in Figure~\ref{fig:mechanism_cartoon}a. The work done by this mechanism upon a given particle is
    \begin{equation} \label{eq:W_direct}
    \frac{W_{\rm direct}}{m_ec^2} = \eta_{\rm rec} \beta_A \big( \omega_{\rm ce} \tau \big) \simeq 0.1 \, \omega_{\rm ce} \tau, 
    \end{equation}
    where~$\eta_{\rm rec} \beta_A \equiv E_{\rm rec}/B_0 \simeq 0.1$ is the reconnection rate normalized by the speed of light. Here~$\beta_A \equiv \sqrt{\sigma_h/(1+\sigma_h)} \simeq \sqrt{\sigma/(1+\sigma)}$ is the dimensionless upstream Alfv\'en speed, $\omega_{\rm ce} \equiv eB_0/m_ec$ is the nominal electron gyrofrequency defined with the upstream magnetic field~$B_0$, and~$\tau$ is the amount of time the particle spends in the X-point \citep{French_2025}. 

    \item A Fermi reflection (a half-cycle of a continual Fermi process) as the particle goes through the relaxation of freshly-reconnected magnetic field-line tension downstream of reconnection X-points. This is due to the alignment of the local curvature drift velocity~$\textbf{v}_c$ with the ideal MHD electric field~$\textbf{E}_m = - \textbf{u} \times \textbf{B}/c$\citep{Fermi1949,Drake2006,Guo2015,Dahlin2014,Qile_2021,Majeski_2023,Qile_2024,French_2025} (Figure~\ref{fig:mechanism_cartoon}b). If modeled as an elastic collision, the particle energy gain  by this mechanism is~\citep{French_2025}: 
    \begin{equation} \label{eq:W_Fermi}
    \frac{W_{\rm Fermi}}{m_ec^2} = \frac{1 + \beta_{\rm Ax}^2}{1 - \beta_{\rm Ax}^2} - 1 = \frac{2\sigma}{1 + \sigma b_g^2}, 
    \end{equation}
    where~$\beta_{\rm Ax} =  \left[\sigma/[1 + \sigma (1 + b_g^2)] \right]^{1/2}$ is the dimensionless in-plane Alfv\'en speed. In a nonrelativistic context, this formula reduces to~$\Delta v = 2 v_{\rm Ax}$ \citep{Drake2006,Dahlin2014}. 

    \item Pickup acceleration in the outflow region, wherein a wandering upstream particle crosses the separatrix and is demagnetized, yielding a sudden increase in perpendicular momentum~$\textbf{p}_\perp$ and acceleration by the reconnection outflow upon entry into the downstream region
    \citep{Drake2009,Sironi2020,French_2023,Chernoglazov2023} (Figure~\ref{fig:mechanism_cartoon}c). The work done by this mechanism is~\citep{Drake2009,French_2023,French_2025}:
    \begin{equation} \label{eq:W_pickup}
    \frac{W_{\rm pickup}}{m_ec^2} \equiv \gamma_{\rm Ax} - \gamma_0 = \sqrt{1 + \frac{\sigma}{1 + \sigma b_g^2}} - \gamma_0,
    \end{equation}
    where~$\gamma_{\rm Ax}$ uses the in-plane Alfv\'en speed~$\beta_{\rm Ax}$ and~$\gamma_0$ is the particle's initial energy.  
\end{enumerate}
These three injection mechanisms act on particles which are making their entry into the downstream from the upstream. Furthermore, all three mechanisms arise naturally from the geometry of reconnection layers, and therefore suggest to exhaust the channels by which a particle may be injected around a reconnection X-point and immediately downstream of it. 

To obtain a more complete description of the injection picture, the contribution of these mechanisms to the total injected particle population has been investigated over a variety of system parameters, including upstream magnetizations \citep{Sironi2022,Guo2022_comment,Totorica_2023,French_2025}, guide-field strengths \citep{Sironi2022,Guo2022_comment,French_2023}. The differences between 2D and 3D have also been investigated directly \citep{Sironi2022, Totorica_2023, French_2025}. 

The time-dependent contributions of each injection mechanism are calculated using the following scheme over a statistically large ensemble of tracer particles. At the timestep at which a tracer particle first has energy~$\gamma > \gamma_{\rm inj}$, it is categorized to a mechanism based on the following criteria:
\begin{equation} \label{eqn:mechanism_categories}
\begin{gathered}
(W_\parallel > W_\perp) \ \& \ (\lvert \textbf{p}_\parallel \rvert > \lvert \textbf{p}_\perp' \rvert) \implies \ \text{$E_{\rm rec}$ acceleration} \\
(W_\perp > W_\parallel) \ \& \ (\lvert \textbf{p}_\parallel \rvert  > \lvert \textbf{p}_\perp' \rvert) \implies \ \text{Fermi kick(s)} \\
(W_\perp > W_\parallel) \ \& \ (\lvert \textbf{p}_\perp' \rvert > \lvert \textbf{p}_\parallel \rvert) \implies \ \text{Pickup process},
\end{gathered}
\end{equation}
with the remaining possibility categorized as ``other" \citep{French_2023,French_2025}. Here $W$ refers to the particle energy gain, $\textbf{p}$ refers to particle momenta in the simulation frame and $\textbf{p}'$ indicates  the momenta in the $E\times B$ drift frame, respectively. Subscripts~$\parallel, \perp$ indicate components relative to the local magnetic field.

\begin{figure*}[htp!]
    \centering
    \includegraphics[width=\linewidth]{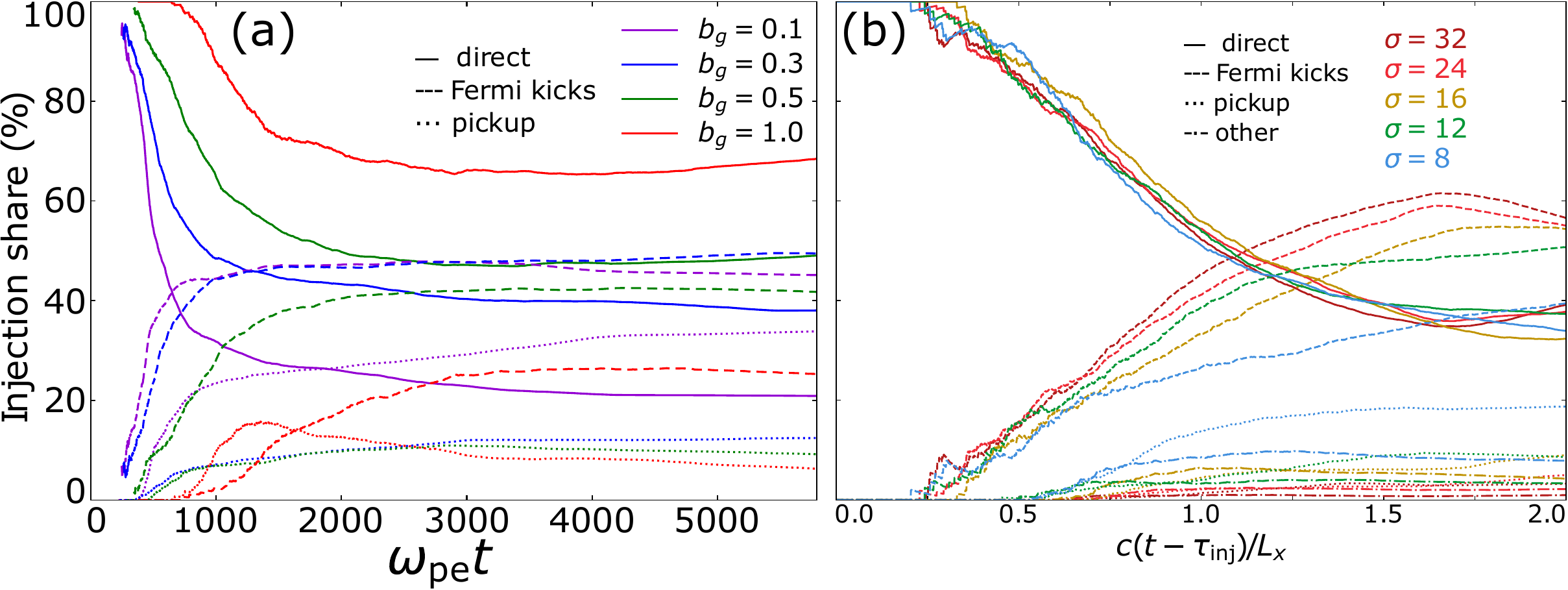}
    \caption{Panel~(a): Time-dependent injection shares of 2D simulations of various guide-field strength, adapted from~\citet{French_2023}. In panel~(a), the cold upstream magnetization is~$\sigma = 50$ and the results are domain-size converged for~$b_g = 0.1, 0.3$, with slightly greater contributions expected from~$W_\perp$ for~$b_g = 0.5, 1.0$ in the large-domain size limit. 
    Panel~(b): Time-dependent injection shares of 3D simulations of various magnetizations~$\sigma$, adapted from~\citet{French_2025}. The time is normalized to when the first particle is injected~$\tau_{\rm inj}$, which is slightly ahead of the reconnection time~$t_{\rm rec}$. In panel~(b), the guide-field strength is~$b_g = 0.3$ and all results are domain-size converged.
    }
    \label{fig:inj_shares}
\end{figure*}

Figure~\ref{fig:inj_shares}a shows the time-dependent contributions of parallel and perpendicular electric fields to the total injected particle population with different guide fields, holding fixed the upstream magnetization $\sigma = 50$ and the upstream temperature $\theta = k_B T/m_e c^2 = 0.25$. The contribution of Fermi kicks and pickup processes are more than other mechanisms for weak guide field case, but direct acceleration becomes more important as guide field becomes stronger. 
In environments where the guide-field is strong (e.g., $b_g \equiv B_g/B_0 \gtrsim 1.0$), works by \citet{Sironi2022, French_2023} have independently concluded that parallel electric fields dominate the injection process. 

Figure~\ref{fig:inj_shares}b shows the contributions of these mechanisms for 3D simulations over a range of upstream magnetizations~$\sigma$ (i.e., the average magnetic energy per particle), holding fixed the guide-field strength $b_g = 0.3$ and the upstream temperature $\theta = 0.3$. Notably, direct acceleration and Fermi kicks remain competitive as~$\sigma$ increases, but the pickup mechanism becomes less efficient to inject particles at sufficiently high~$\sigma$.

There have been different conceptions about how to evaluate the relative importance of various injection mechanisms. In~\citet{French_2023,French_2025}, the \textit{injection shares}---i.e., the fractional contribution of each mechanism to the total injected particle population--- were chosen to reflect the relevance of each mechanism. In contrast, works by~\citet{Totorica_2023} and~\citet{Gupta_2025} consider the NTPA correlations evaluated at the a small fraction of highest accelerated particle to reflect the importance of the mechanism they study. However, as mentioned in~\citet{Kilian2020,French_2025}, this latter approach may produce a significant systematic bias towards particles that are injected at earlier times and accelerated to highest energy, and these particles tend to be injected by direct acceleration. It is worth noting that the primary injection mechanism should be able to explain how the majority of injected particles are accelerated to the injection energy~$\varepsilon_{\rm inj}$. In addition, while some particles can get initial boost from diffusion regions and reach high-energy in the beginning, in a large-scale reconnection layer they will mix with other injected particles if some second-order acceleration exists, as demonstrated in \citet{Li2023}. 

Nevertheless, the injection shares can be derived from Figure~2b of~\citet{Totorica_2023} by evaluating each line at~$\epsilon_{\rm final} = \epsilon^*$. For~$\sigma = 50, \, b_g = 0.125$, this then yields nonideal electric field~$E_n$ contributing~$\sim 30\%$ to the injected particle population for the energy threshold~$\epsilon^* = \sigma/4$, in rough agreement with \citet{French_2023} which found~$E_\parallel$ to contribute~$\sim 20\%$. Likewise, the injection shares can be read off Figure~3 of~\citet{Gupta_2025}, which implies that~$E>B$ electric fields inject~$\sim 5\%$ of particles, in rough agreement with~\citet{Guo2022_comment}. These studies reached different conclusions mainly because their different focuses and approaches in evaluating the importance of injection mechanisms.

\begin{figure}[htp!]
    \centering
    \includegraphics[width = 0.7\textwidth]{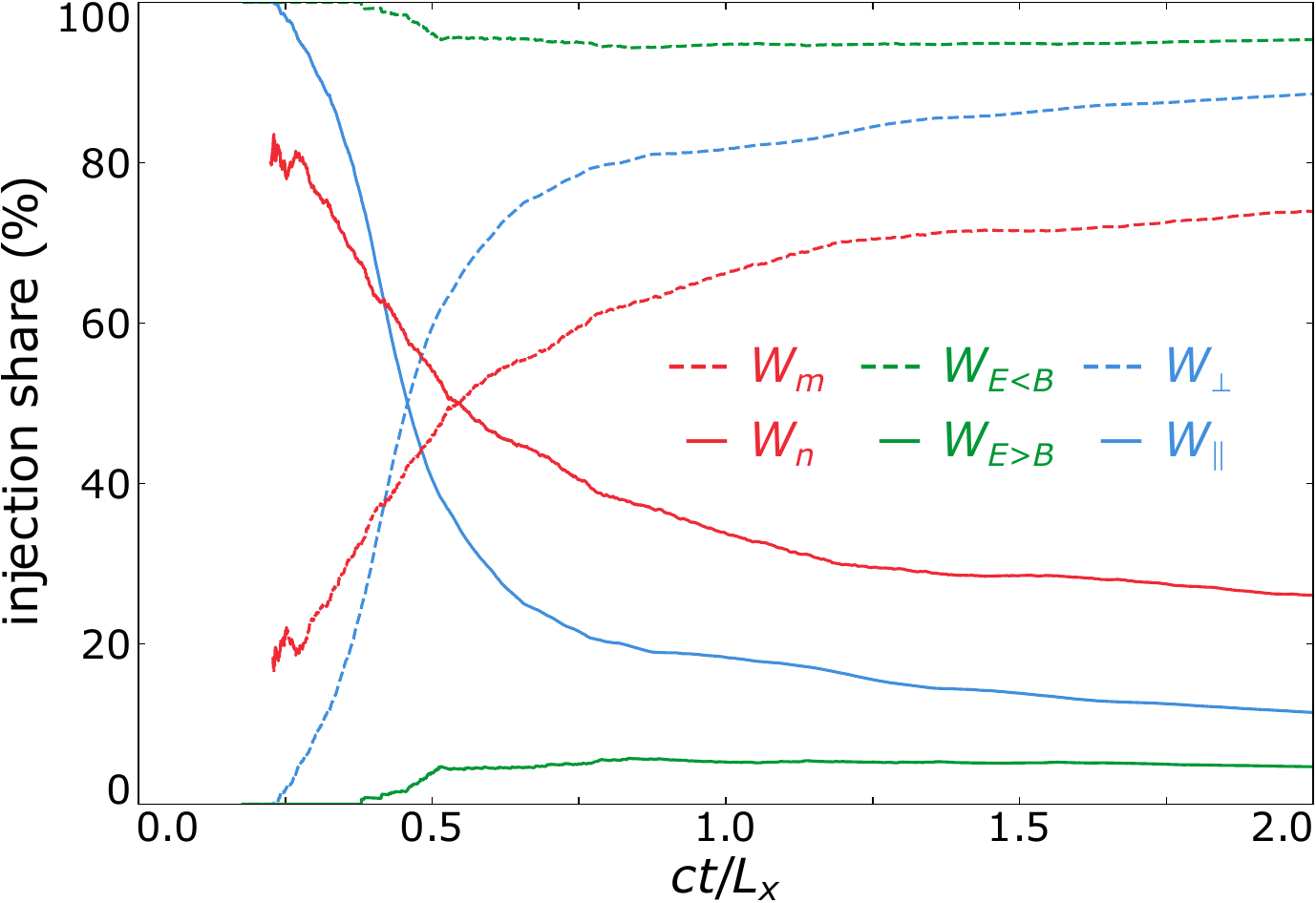}
    \caption{Comparison between three injection dichotomies, using~$b_g = 0$ and~$\sigma = 50$. Red: Fraction of particles that reached the injection kinetic energy~$\varepsilon_{\rm inj}/m_ec^2 = 10$ that gained most of their energy from (i.e., ``injected by") the non-ideal electric field~$E_n$ (solid) versus the ideal electric field (dashed). Green: Fraction of particles injected by~$E>B$ regions (solid) versus~$E<B$ regions (dashed). Blue: Fraction of particles injected by parallel electric fields (solid) versus perpendicular electric fields (dashed).}
    % \caption{Time-evolved contributions from each injection mechanism to the total injected particle population. Each panel contains a different injection dichotomy. Panel~(a):  }
    \label{fig:Wn_vs_Wegb}
\end{figure}

While many studies decompose the electric field into parallel $\textbf{E}_\parallel$ and perpendicular $\textbf{E}_\perp$ components to evaluate the acceleration mechanisms. Several studies have distinguished the full non-ideal electric field and motional electric field based on the generalized Ohm's law  \citep{Guo2019,Totorica_2023}. The non-ideal MHD electric field~$\textbf{E}_n \equiv \textbf{E} + \textbf{u} \times \textbf{B}/c$ that generally operates on scales~$\ell \lesssim c/\omega_{\rm pe}$, and the ideal (i.e., motional) electric field~$\textbf{E}_m \equiv -\textbf{u} \times \textbf{B}/c$. The $\textbf{E}_\parallel$-$\textbf{E}_\perp$ way can be computationally easier and faster as it does not need to collect the particle moments, and can single out the direct acceleration parallel to the magnetic field. However, $\textbf{E}_n$-$\textbf{E}_m$ decomposition may work better when the guide field is nearly zero. While earlier studies reach the same conclusion for the importance of $\textbf{E}_\perp$ and $\textbf{E}_m$ for weak guide field cases, a recent paper by~\citet{Totorica_2023} disputes this conclusion. In Figure~\ref{fig:Wn_vs_Wegb}, we compare these injection dichotomies by plotting the time-dependent contributions of~$W_n$ vs~$W_m$ to the total injected particle population (red) and~$W_{\rm E>B}$ vs $W_{\rm E<B}$ to the total injected particle population (green), where~$W_{\rm E>B}$ ($W_{\rm E<B}$) is work done by the electric field in~$E>B$ ($E<B$) regions. In this run, the upstream magnetization~$\sigma = B_0^2/4\pi n_0m_ec^2 = 50$ (where $n_0 = n_e + n_i$ is the total particle number density) and we assume that the injection kinetic energy~$\varepsilon_{\rm inj}/m_ec^2 = 10$ for simplicity and proximity to measured values \citep{French_2023}. These results still show that perpendicular electric fields / motional electric fields are more important for weak guide field than parallel electric field / non-ideal electric field. Meanwhile, the contribution of $E>B$ region is nearly ignorable. %reaffirm the Comment by~\citet{Guo2022_comment}, demonstrating that~$E>B$ regions are only responsible for injecting~$\sim 1/20$ of the injected particle population. The results also demonstrate that non-ideal electric fields, while initially dominant, are only responsible for injecting~$\sim 25\%$ of the aggregate injected particle population after~$\sim 2 \,L_x/c$ have passed---in rough agreement with the~$b_g = 0.1$ case shown in Figure~\ref{fig:French2023_inj_shares}.

%\subsection{Finite guide-field strength: The $E_\parallel$ vs $E_\perp$ dichotomy} 

%To simplify the dichotomy discussed above, one strategy is to identify the non-ideal electric field~$\textbf{E}_n$ with the electric field parallel to the local magnetic field~$\textbf{E}_\parallel \equiv (\textbf{E} \cdot \textbf{B})\textbf{B}/\lvert \textbf{B} \rvert^2$ and the motional electric field~$\textbf{E}_m$ with the perpendicular electric field~$\textbf{E}_\perp \equiv \textbf{E} - \textbf{E}_\parallel$ ~\citep{Guo2019,Ball2019,Kilian2020,Sironi2022,French_2023,Guo2023_comment,French_2025}. These assignments considerably simplify the subsequent tracer particle analysis, as the flow velocity~$\textbf{u}$ does not need to be tracked. At this stage, the question of the dominant injection mechanism is reduced to comparing the work done by each component (e.g., $W_\parallel(t) \equiv \int_0^t \textbf{v}(s) \cdot \textbf{E}_\parallel(s) \,ds$, $W_\perp(t) \equiv \int_0^t \textbf{v}(s) \cdot \textbf{E}_\perp(s) \,ds$) over a statistically large ensemble of tracer particles. 

% \begin{figure}[htp!]
%     \centering
%     \includegraphics[width=0.5\textwidth]{figures/injection_Wn_vs_bg.pdf}
%     \caption{Correspondence between the injection shares for~$W_n$ for~$b_g \lesssim \eta_{\rm rec}$ to~$W_\parallel$ for~$b_g \gtrsim \eta_{\rm rec}$. % \OF{If we want~$W_{E>B}$ or~$W_{E<B}$ can also be shown, displaying how it doesn't correspond with~$W_n$ or~$W_m$ (in accordance with \citet{Guo2022_comment} and \citet{Totorica_2023}).}
%     }
% \label{fig:wn_wpara_correspondence}
% \end{figure}
% \OF{Figure: For~$b_g = 0.1$, show injection shares of $W_n$ vs $W_m$, $W_\parallel$ vs $W_\perp$. The shares are within~$\sim 5\%$ error.}

\subsection{Injection Efficiency and Energy Partition between Thermals and Nonthermals}

\begin{figure*}[htp!]
    \centering
    \includegraphics[width= \linewidth]{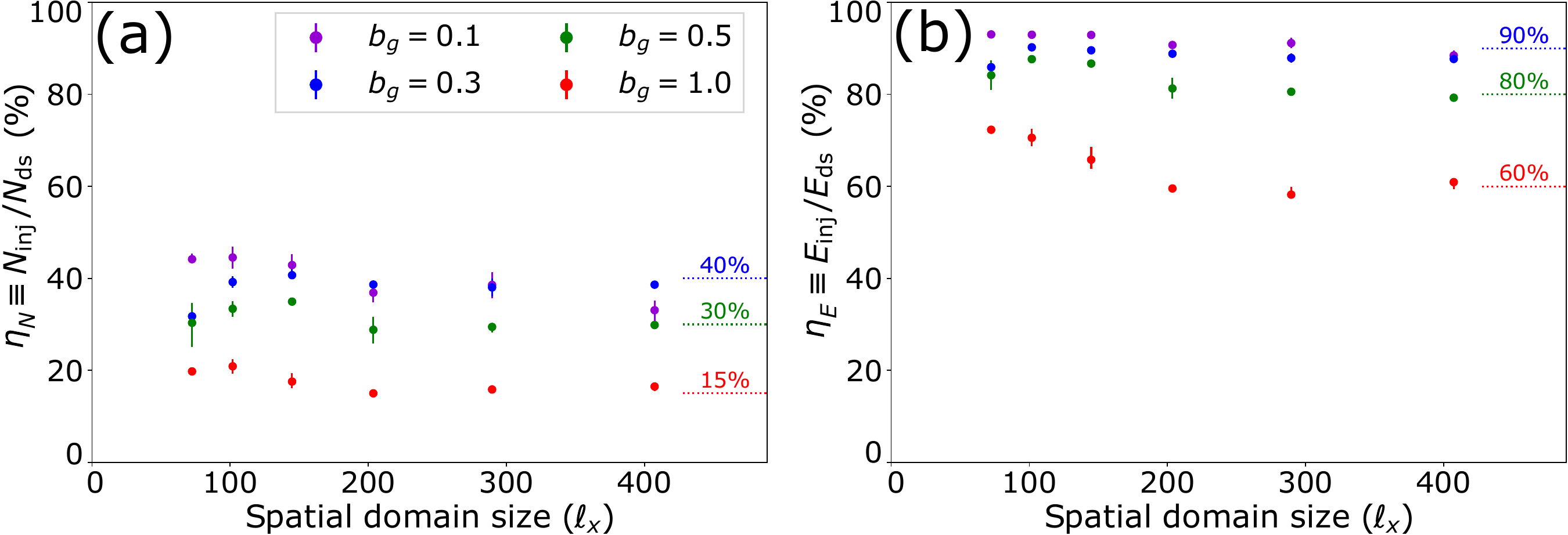}
    \caption{Efficiencies of (a) injection and (b) energy at final times over various guide-field strengths~$b_g$ and domain sizes~$\ell_x$, adapted from~\citet{French_2023}. }
    \label{fig:efficiencies}
\end{figure*}

As we gain physics understanding of particle injection during magnetic reconnection, it allows us to quantify the efficiency of relativistic magnetic reconnection in generating nonthermal particles. Thus far, two similar definitions have been employed to quantify the efficiency. ~\citet{Hoshino_2022,Arnold2021} fit the distribution as a Kappa function and obtain the nonthermal population by removing the corresponding Maxwellian distribution \citep{Oka2022}.
This is well-suited for nonthermal particle spectra with short dynamic ranges, such as those with transrelativistic hot upstream magnetizations~$\sigma_h = B_0^2/4\pi h = \sigma/\Gamma_{\rm th} \sim 1$, where~$\Gamma_{\rm th} \equiv K_3(\theta^{-1})/K_2(\theta^{-1})$ is the ``thermal Lorentz factor", i.e., the average energy of electrons in a Maxwell-J\"uttner distribution of temperature~$\theta \equiv k_B T/m_e c^2$. 

% The formal definition is
% \begin{align} \label{eqn:kappa_inj_efficiency}
%     \eta_N \equiv N_{\kappa}/N_{\kappa + \rm M}, \\
%     N_{\kappa} \equiv \int_1^\infty f_{\kappa}(\gamma) \,d\gamma, \, \, N_{\kappa + \rm M} \equiv \int_1^\infty f_{\kappa + \rm M}(\gamma) \,d\gamma. \nonumber
% \end{align}

 \citet{French_2023} quantify nonthermal power-law spectra, in which the number efficiency is defined by the fraction of the downstream particle population that has undergone injection into the power-law spectrum, i.e., $\eta_N \equiv N_{\rm inj}/N_{\rm ds}$, where
\begin{align} \label{eqn:PL_inj_efficiency}
    N_{\rm inj} \equiv \int_{\varepsilon_{\rm inj}}^\infty f_{\rm ds}(\varepsilon) \,d\varepsilon, \, \, N_{\rm ds} \equiv \int_0^\infty f_{\rm ds}(\varepsilon) \,d\varepsilon. \nonumber 
\end{align}

Meanwhile, energy efficiency can be defined as well, which is the fraction of downstream particle energy contained by injected particles, i.e., $\eta_E \equiv E_{\rm inj}/E_{\rm ds}$, where
\begin{equation} \label{equation:e_inj}
    E_{\rm inj} \equiv \int_{\varepsilon_{\rm inj}}^\infty \varepsilon f_{\rm ds}(\varepsilon) \,d\varepsilon, \, E_{\rm ds} \equiv \int_0^\infty \varepsilon f_{\rm ds}(\varepsilon) \,d\varepsilon.
\end{equation}
Figure~\ref{fig:efficiencies} shows that for relativistic reconnection with $\sigma_c = 50$, the density injection efficiency can approach~$40\%$ for weak guide field and decrease with guide field, reaching ~$\sim 15\%$ for $b_g = 1$. The energy efficiency can even reach $90\%$ in the weak guide field case and becomes $\sim 60\%$ when the guide field is comparable to the reconnecting magnetic field. 

It is worthwhile to note that when multiple components present in the spectra, it becomes difficult to clearly distinguish the nonthermal component or fit the spectrum for obtaining the efficiency \citep{Hoshino_2024}. Clearer results can be obtained for local spectra or downstream spectra \citep{French_2023,Sironi2020}

\section{Particle Injection in Nonrelativistic and transrelativistic magnetic reconnection} \label{sec:nonrel_recon}

Substantial nonthermal particle acceleration (NTPA) has also been found in simulations of nonrelativistic ($\sigma \ll 1$) and transrelativistic ($\sigma \sim 1$) magnetic reconnection (see recent reviews by \citet{Li2021,Oka2023,Guo2024,Drake2025}). Studying these regimes are of particular importance to understanding reconnection of the heliospheric current sheet \citep{Phan_2022,Desai_2022,Murtas2024}, Earth's magnetotail \citep{Ergun_2020,Oka2023}, and solar flares \citep{Chen2020,Li2022}, as well as steady and flaring emissions from accretion disks around supermassive black holes \citep{Ripperda_2020,Nathanail_2022}.

\subsection{Nonrelativistic Reconnection}

\begin{figure*}[htp!]
    \centering
    \includegraphics[width = \textwidth]{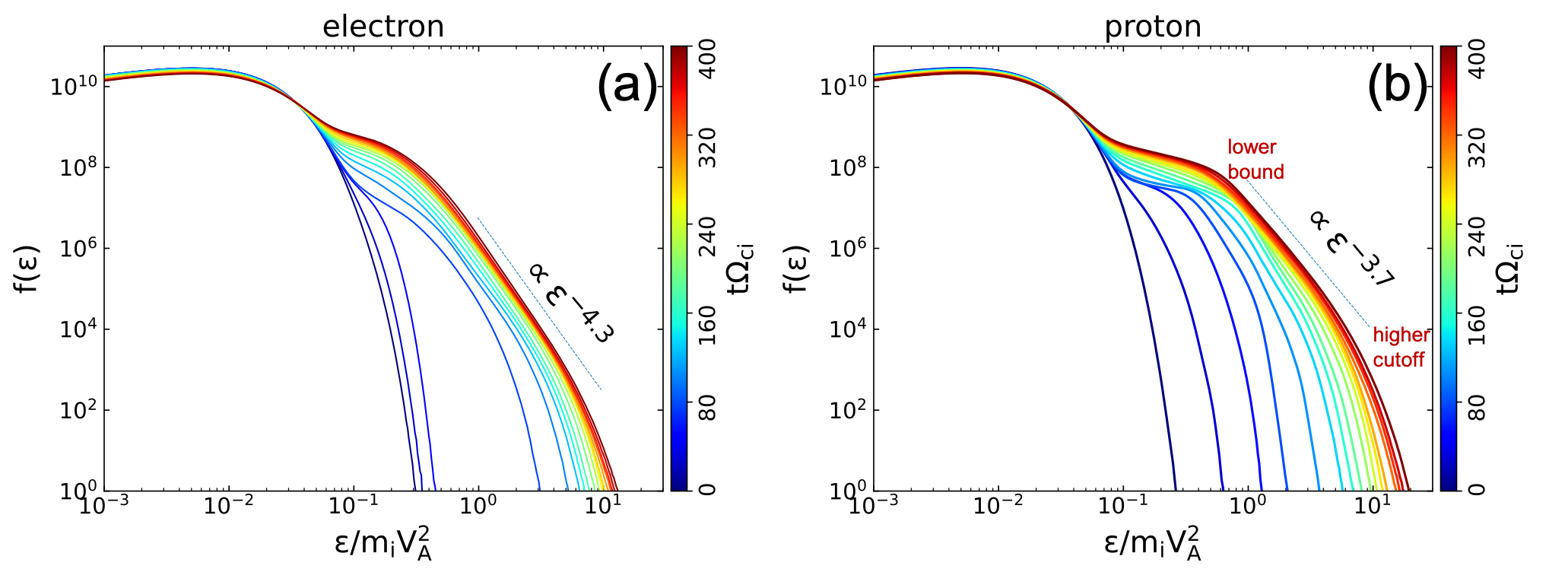}
    \caption{Evolution of energy spectra for electrons (a) and protons (b) in a $L_x=300 d_i$ nonrelativistic simulation. The spectral shapes at high energy resembles power-laws with spectral indices around $4$. The lower bound of the power-law is featured by a shoulder structure, indicating the injection process. es. [Adapted from
\citet{Zhang2021}]}
    \label{fig:Zhang2021}
\end{figure*}

\begin{figure}
    \centering
    \includegraphics[width=0.7\textwidth]{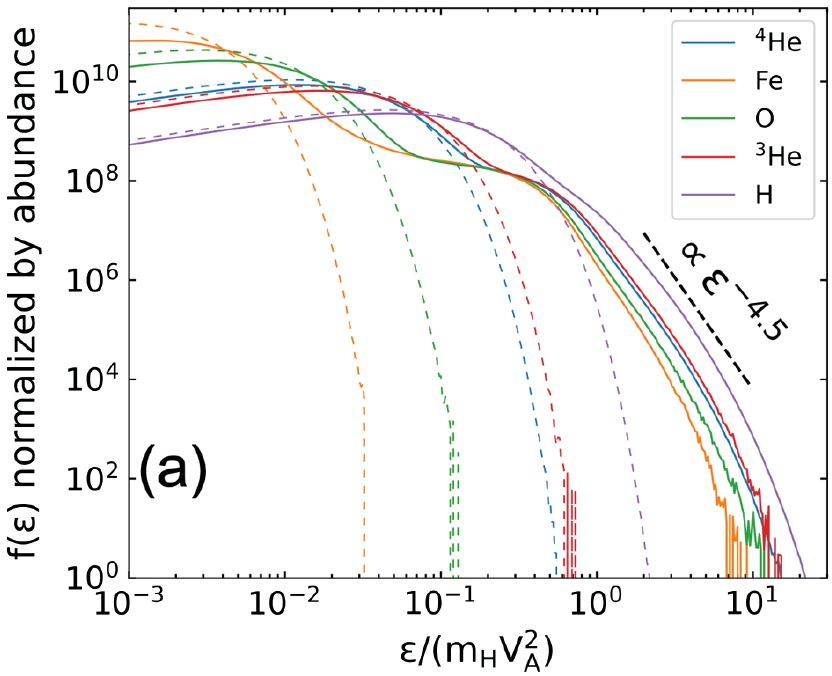}
    \caption{Energy spectra of multi-species ions adapted from~\citet{Zhang2024a}. Final particle-number spectra (solid lines) normalized by the abundance ratio to Fe against energy per nucleon~$\varepsilon$. The initial spectra (dashed lines) are also shown for reference. Each ion species develops a power-law spectrum with a shoulder structure at the lower-bound of the nonthermal component.}
    \label{fig:qile_2024_fig2a}
\end{figure}

Recent 3D kinetic simulations of nonrelativistic reconnection have obtained clear power-law energy distributions, which allows studies of the injection processes \citep{Li2019b,Zhang2021,Zhang2024a}. The energy spectra for different species (electrons, protons, and heavier ions)  show clear ``shoulder'' structures around $\sim m_p v_A^2$, where $m_p$ is the proton mass and $v_A$ is the upstream Alfven speed. 
The shoulder marks the injection energy and determines the energy partition between thermal and nonthermals for different species. Taking Figure \ref{fig:Zhang2021} as an example, \citet{Qile_2021} showed the low-energy bounds of the power-law in the low beta and low guide field regime is $\varepsilon_{lp} \sim 0.5m_p v_A^2$ for protons and $\varepsilon_{le} \sim 0.2m_p v_A^2$ for electrons. $20 \%$ of particles and $\sim 50 \%$ energy of each species can be nonthermal above the low-energy bounds. Correspondingly, nonthermal protons take about twice as much energy as electrons due to the more efficient injection process. 
 ~\citet{Qile_2024} used 3D hybrid (fluid electrons, kinetic ions) simulations to demonstrate for the first time that multispecies ions from Helium to Iron can experience efficient acceleration into a nonthermal power-law spectrum. Regardless of the ion species, the authors find energy spectra with similar power-law indices, suggesting a universal mechanism that governs particle acceleration within the power-law spectrum (Figure~\ref{fig:qile_2024_fig2a}). All ion species develop the shoulder structure indicative to the injection process. 

 The generation of the power-law spectra immediately raises the question of how these ions and electrons are injected from a thermal upstream into the power-law. \citet{Zhang2021} addressed this question by considering the work done to an incoming upstream particle by a single Fermi reflection \citep{Fermi1949,Drake2006,Majeski_2023} off an Alfv\'enic outflow at a reconnection exhaust, i.e.
$$ W_{\rm Fermi} = \frac{1}{2} m_H ( 2v_{\rm out} )^2, $$
where~$v_{\rm out}$ is the outflow velocity and~$m_H$ is the proton mass. For this nonrelativistic reconnection case $v_{\rm out} \sim 0.5 V_A$ and $W_{\rm Fermi}\sim 0.5 m_p V_A^2 $, agreeing with the proton shoulder energy seen in Figure \ref{fig:Zhang2021}. Electron injection for proton-electron plasmas is more complicated and discussed below.  %To estimate the injection kinetic energy~$\varepsilon_{\rm inj}$, the authors sum~$W_{\rm Fermi}$ with the initial particle kinetic energy~$\varepsilon_0 = \frac{1}{2} m_H (2v_{\rm th})^2$, where~$2v_{\rm th} = 2\sqrt{T/M}$ is the velocity of particles of mass~$M$ on the warmer side of the initial Maxwellian spectrum of temperature~$T$. Comparing the theoretically-estimated~$\varepsilon_{\rm inj} = W_{\rm Fermi} + \varepsilon_0$ for different species to the lower bound of the power-law spectra in Figure~\ref{fig:Zhang2021} that marks the injection energy, the authors discover a direct correspondence.

Interestingly, while Fermi reflections appear capable of injecting all ion species, the injections are influenced by their initial thermal velocities
$V_{th}=\sqrt{T_0/M}$ (lower for heavier ions). A particle around the initial thermal velocity will get kicked by the exhaust and gain twice of the outflow speed. Estimating the injection energy per nucleon from a single Fermi reflection 
\begin{equation}
    \varepsilon_{\rm inj}\sim0.5m_H(2V_{\rm th}+2V_{\rm out})^2=2m_H(V_{\rm th}+V_{\rm out})^2. \label{inj_en}
\end{equation}
 This theoretical estimate agrees reasonably well with the shoulder energy in Figure 9.
 
\begin{figure*}[htp!]
    \centering
    \includegraphics[width = \textwidth]{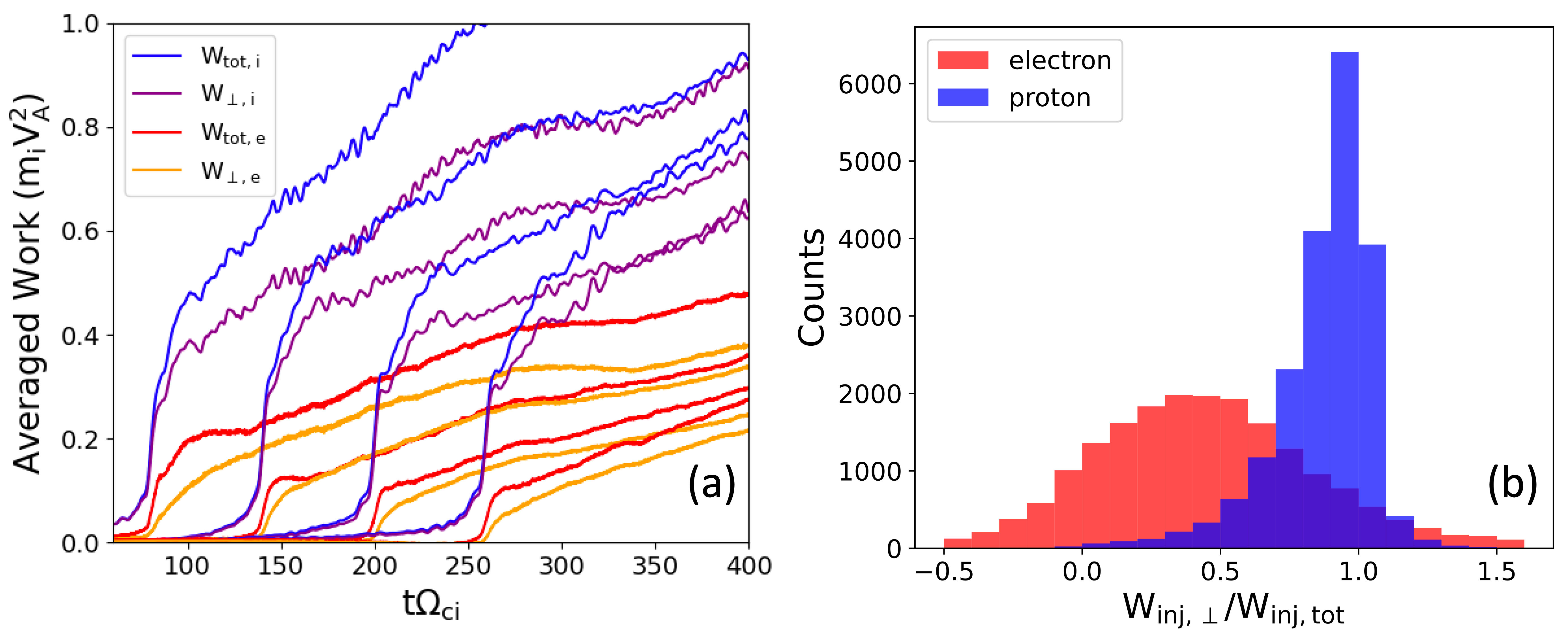}
    \caption{(a) Evolution of work done by total and the perpendicular component of electric fields for protons and electrons, averaged over different generations of injected particles. (b) histogram of the perpendicular-work contribution fraction for injection for both species. [Adapted from
\citet{Zhang2024ApJ}, reproduced by permission of the AAS]  \label{fig4}}
    % \label{fig:mechanism_cartoon}
\end{figure*}

Now we discuss the relative contribution of perpendicular and parallel electric fields during the particle injection process for protons and electrons. \citet{Zhang2024ApJ}  show the evolution of the total work done and its perpendicular-electric-field component, averaged within each different generation of electrons and protons entering the reconnection region and starting acceleration at different time, as illustrated in Figure \ref{fig4}(a). Particles are included in a generation if the final energies are
above the spectral low-energy bound and if the starting time of energization
is within an $\Omega_{ci} \Delta t=5$ interval. It shows that the protons are mostly injected by the perpendicular electric field, while the electrons are only half injected by it. They also show a histogram of injection contribution percentage (Figure \ref{fig4}(b)), suggesting a similar conclusion: the proton perpendicular contribution peaks around 100\% while electrons' peaks around 50\%. 
 While both electrons and protons are injected at reconnection exhausts, protons are primarily injected by perpendicular electric fields through Fermi reflections and electrons are injected by a combination of perpendicular and parallel electric fields.

\begin{figure*}[htp!]
    \centering
    \includegraphics[width = \textwidth]{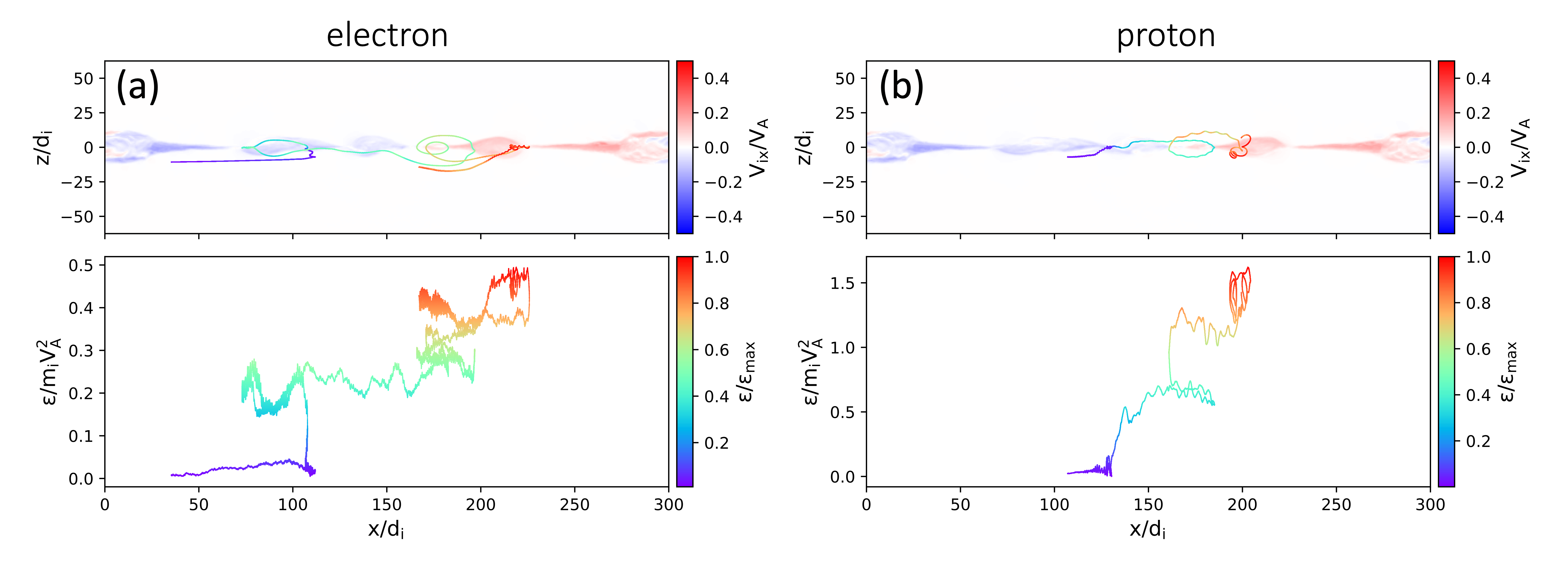}
    \caption{Representative particle trajectories in a nonrelativistic simulation demonstrating the injection process for both species, colored by the particle energy $\varepsilon$ normalized by the maximum energy $\varepsilon_{max}$. The top panels show proton velocity in $x$ ($V_{ix}$) as backgrounds around the injection time of the particles (a 2D $x-z$ slice at the $y$ location where the particle is injected. [Adapted from
\citet{Zhang2024ApJ}, reproduced by permission of the AAS]}
     \label{fig5}
\end{figure*}

 We also show two representative particles (an electron and a proton) to demonstrate their injection process at reconnection exhausts. Figure \ref{fig5}(a-b) (adapted from \citet{Zhang2024ApJ}) shows the background of $V_{ix}$ around the injection time of the particles, with the particle trajectories (colored by energy) overlaid. When particles for the first time cross an exhaust from upstream, they get boosted to the injection energy (around $0.2m_iV_A^2$ for electrons and $0.5m_iV_A^2$ for protons). 
%As shown in the electric-field-component contribution above, the protons are injected by a Fermi reflection at the exhaust, and electrons by a combination of Fermi reflection and parallel electric fields. 
After injection, the particles wander elsewhere and get further Fermi acceleration.

% \citet{Qile_2021} illustrates the importance of the flux-rope-kink instability (FRKI) in transporting particles to acceleration zones. 
\subsection{Transrelativistic regime of reconnection}
Recent studies have also studied the so-called trans-relativistic regime, where $\sigma_i < 1$ but  $\sigma_e \gg 1$, which can lead to strong electron energization \citep{Werner2018,Ball2018,Kilian2020,Li2023}. The transrelativistic regime is also a bridge for connecting the highly relativistic regime ($p \gtrsim 1.5$) with the nonrelativistic reconnection studies ($p \sim 4$) \citep{Dahlin2014,Li2021,Li2019a,Zhang2021}. 

\begin{figure*}[htp!]
    \centering
    \includegraphics[width = \textwidth]{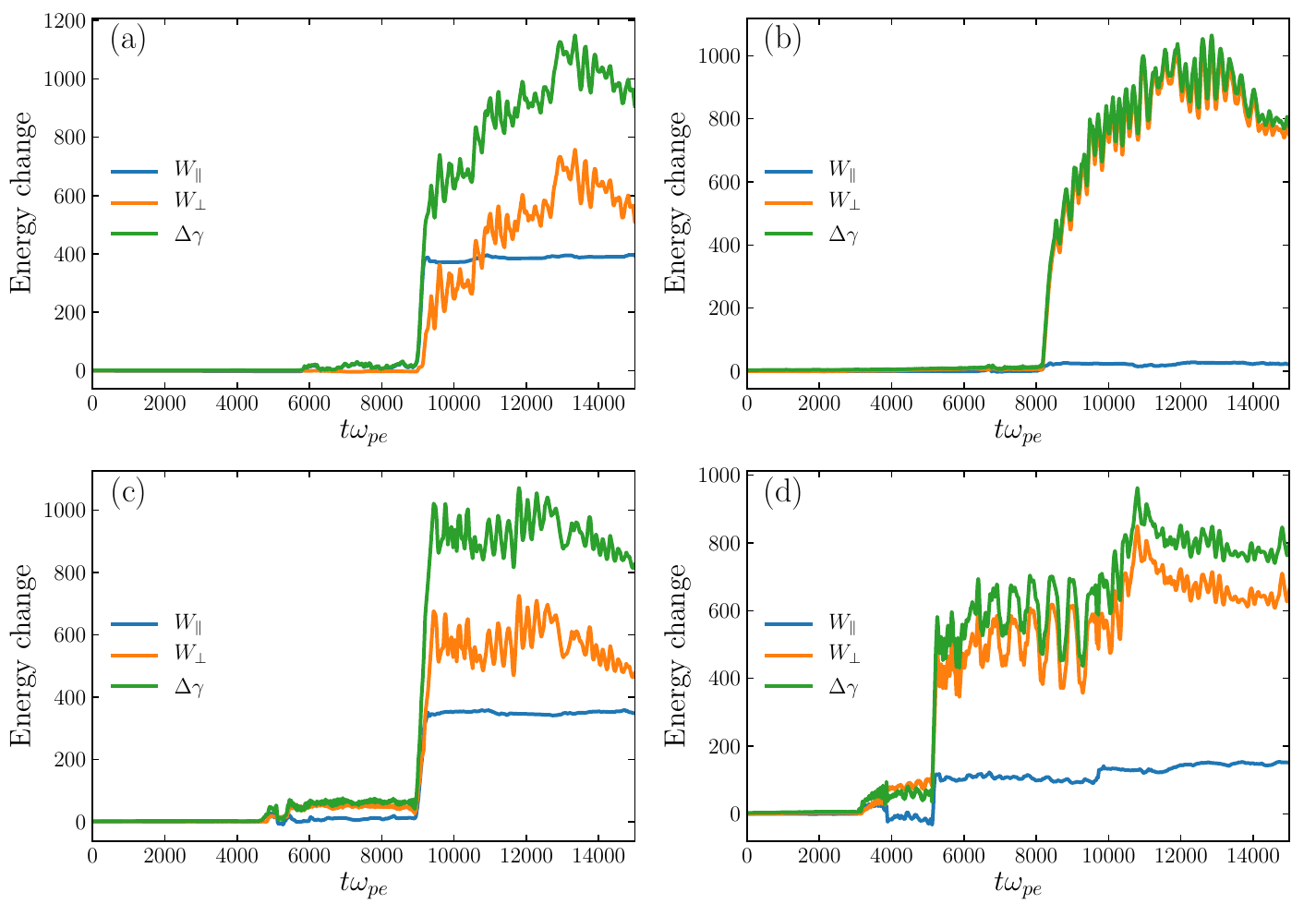}
    \caption{Four electron trajectories in a trans-relativistic reconnection simulation with $\beta = 3.3\times 10^{-3}$, $\sigma_i = 0.3$, and $\sigma_e = 552.5$ (adapted from \citet{Kilian2020}). a) The top left trajectory illustrates the case where a particle is first injected due to $W_\parallel$ and then gains further energy due to $W_\perp$. b) The second trajectory illustrates that $W_\perp$ can directly inject particles with negligible contribution from $W_\parallel$. c) The third trajectory illustrates that both components of the electric field can act simultaneously during injection. d) The fourth trajectory illustrates that for many particles classification of the injection mechanism is not straight forward.}
     \label{trans-rel-tracers}
\end{figure*}

%There have been several studies on the injection problem with energization up until the injection energy $\gamma_{inj}$. While \citet{Ball2019} focused on the work done by the parallel electric field ($W_\parallel$), \citet{Kilian2020} studied the roles of both parallel and perpendicular electric fields. They showed that both parallel and perpendicular electric fields play a role, and perpendicular electric fields become more important for particle injection, especially when the simulation domain becomes large. 

\begin{figure*}[htp!]
    \centering
    \includegraphics[width = \textwidth]{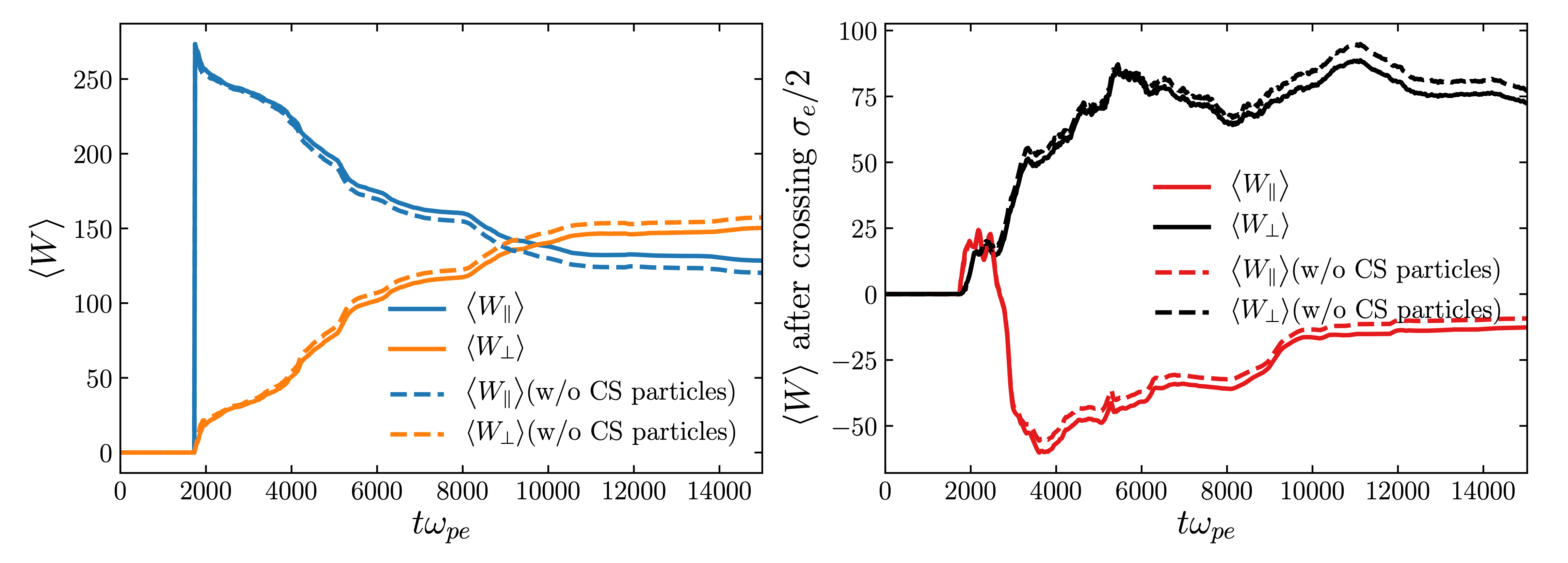}
    \caption{Contributions of the work done by the parallel and perpendicular electric fields to particle injection. Left panel: $W_\parallel$ and $W_\perp$ are averaged over all tracer particles that have crossed the injection threshold by time $t$. Right panel: Average energy gain due to parallel and perpendicular electric field of all particles that have already crossed the threshold $\gamma > \sigma /2$ as a function of time. After a very short initial time, energy gain is dominantly due to the perpendicular electric field while the parallel field removes energy from the particles. [Adapted from
\citet{Kilian2020}, reproduced by permission of the AAS]}
     \label{W-transrel}
\end{figure*}

\citet{Ball2018} discussed the acceleration of highest-energy particles by tracing their trajectories, and claim that X-point acceleration are important for accelerating electrons in the low-$\beta$ regime. However, they only analyzed and showed several trajectories with highest energies. This kind of approach can potentially be biased and cannot reach a full picture. Seen the reconnection acceleration studies in relativistic and nonrelativistic regimes, it is not surprising that Fermi-like acceleration can be important in the trans-relativistic regime. Figure \ref{trans-rel-tracers} shows several particle trajectories on their energies as a function of time with acceleration contribution from both parallel and perpendicular electric fields (adapted from \citet{Kilian2020}), where the reconnection layer is well in the low-$\beta$ transrelativistic regime ($\beta = 3.3\times 10^{-3}$, $\sigma_i = 0.3$, and $\sigma_e = 552.5$). They clearly show that the initial acceleration can be contributed to both parallel and perpendicular electric fields. 
\citet{Ball2019} implemented a technique to identify electrons injected near the X-points, and associated the acceleration in the vicinity of X-points as the acceleration by the parallel electric field. However, there can be several different acceleration mechanisms operating close to the X-points and therefore this kind of analysis can be problematic \citep{French_2023}.  A physics-robust method, as discussed in Section 2, is to carefully analyze the trajectories of a statistically meaningful sample of particles with their local electric and magnetic fields tracked at every time step. The left panel of Figure \ref{W-transrel} shows that the mean energy gain before the electron acceleration of $\gamma \sim \gamma_{\rm inj} \sim \sigma_e/2$. While the initial acceleration is more associated with the parallel electric field, perpendicular electric field catches up quickly, and eventually exceeds the contribution of parallel electric field. After injection, the contribution of perpendicular electric field (right panel) completely dominates the energy gain, whereas the contribution from parallel electric field is nearly negligible.

\begin{figure*}[htp!]
    \centering
    \includegraphics[width = 0.9\textwidth]{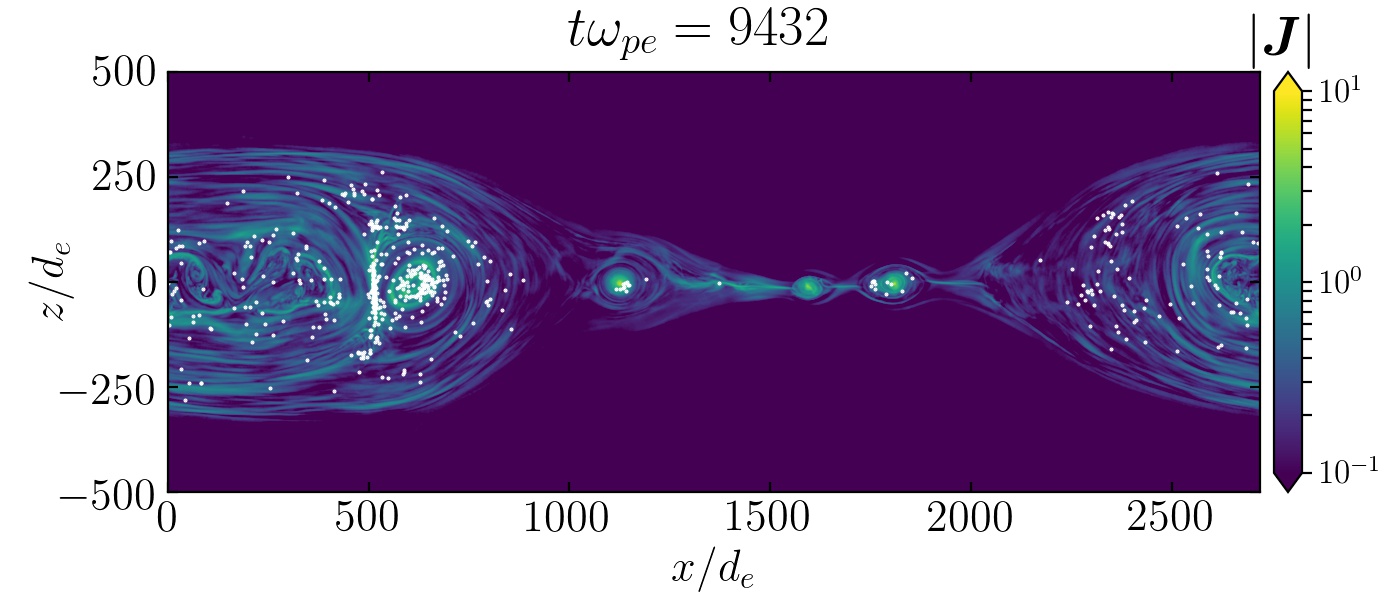}
    \caption{Current density and particle location at $t = 9432\,\omega_{\rm pe}^{-1}$ for particles that cross $\gamma_{\rm inj} = \sigma_e / 2$ around that time $9393 < t\,\omega_{\rm pe} < 9471$. [Adapted from
\citet{Kilian2020}, reproduced by permission of the AAS]}
     \label{Inj-location}
\end{figure*}

Figure 
\ref{Inj-location} shows the current density and location of tracer particles around the time that they are accelerated to $\gamma_{\rm inj} = \sigma_e/2$. This shows that the location of particle injection can be quite broad, with contributions from X-lines, islands and exhaust regions. This again shows that the particle injection can be via different mechanisms rather than limited to X-points.

\begin{figure*}[htp!]
    \centering
    \includegraphics[width = \textwidth]{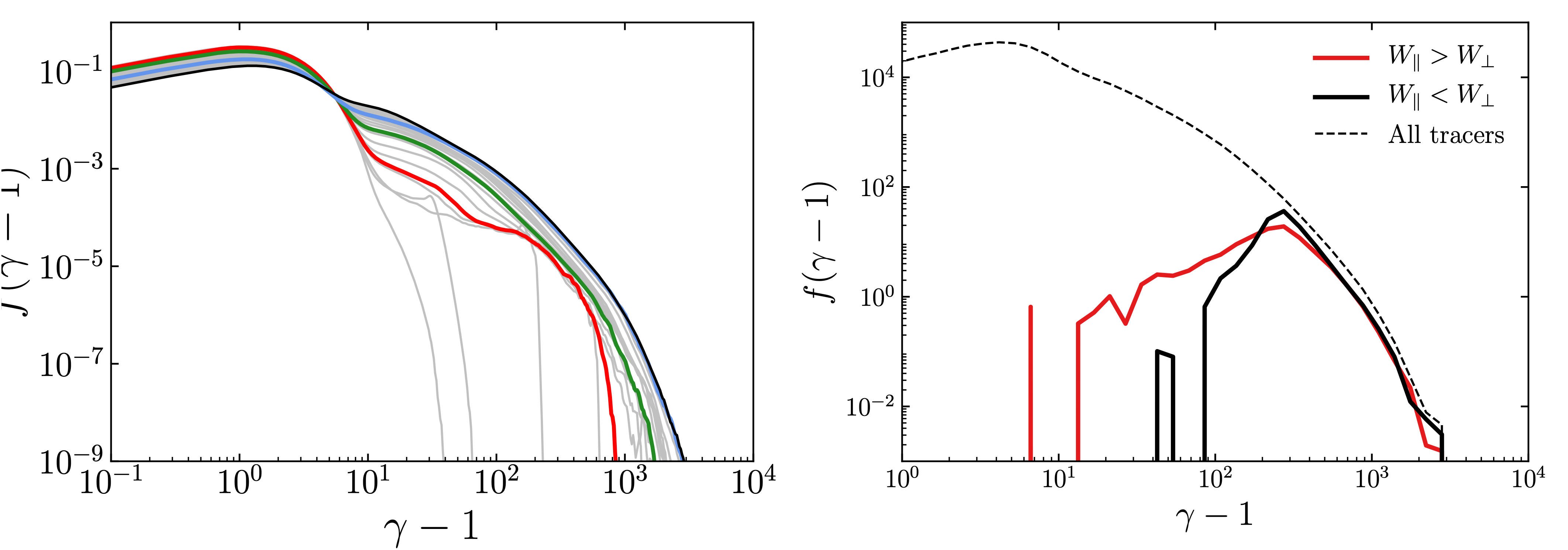}
    \caption{Left: Particle spectra of all tracer electrons at different times. The light gray lines are separated by approximately $780\,\omega_{\rm pe}^{-1}$. The black line shows the energy spectrum at the end of the simulation. Right: Particle spectrum of electrons that did not start in the current sheet at the end of the simulation, divided according to the most work done by the parallel or perpendicular field up to the moment when they cross $\gamma_{\rm inj} = \sigma_e / 2$. The two population shows very similar spectra. [Adapted from
\citet{Kilian2020}, reproduced by permission of the AAS]}
     \label{Spectra-trans-rel}
\end{figure*}

During the injection and prolonged acceleration phases, the acceleration to high-energy is marked by the formation of power-law distributions. Figure \ref{Spectra-trans-rel}(a) shows the evolution of energy spectrum as a function of time, with a clear power-law distribution at high energy. Figure \ref{Spectra-trans-rel}(b) separates the components with dominant contribution from acceleration by parallel electric field and perpendicular electric field. Note that the spectrum of these two group of particles are nearly the same, meaning that the later stage forms the acceleration of particles, and the injection process does not strongly influence the spectral indices.

\section{Magnetically-dominated turbulence} \label{sec:turb}

Relativistic turbulence in the magnetically dominated regime has shown impressive results on nonthermal particle acceleration \citep{Zhdankin_2017,Comisso2018,Comisso2019}. This is especially true for the case with large amplitude fluctuation cases, presumably due to the compressible nature of turbulence in this regime.
Fully kinetic PIC simulations of relativistic plasma turbulence have found nonthermal particle power-law spectra to be a generic outcome \citep{Comisso2018,Comisso2019}. The resulting power-law hardens with larger upstream magnetizations~$\sigma$ and stronger turbulent fluctuations, reaching efficient acceleration~$p \lesssim 2$. Magnetic turbulence has been considered to have intimate connection with magnetic reconnection, and potentially particle acceleration in these processes have close relations. 

\begin{figure}
    \centering
    \includegraphics[width=0.7\textwidth]{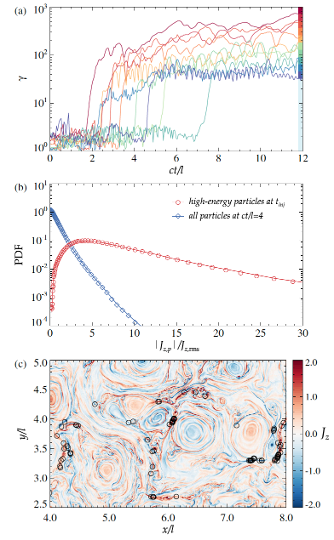}
    \caption{(a) Evolution of the Lorentz factor for 10 representative particles ending up in different high energy bins. (b) PDFs of current density experienced by the injected particles at their injection time $t_{inj}$ (red) and by all the tracked particles (blue). (c) Zoom of the current density in the out of plane direction~$J_z$ with circles indicating the positions of the particles that are injected around the time [Adapted from \citet{Comisso2018} ]. }
    \label{fig:Comisso}
\end{figure}

There have been several studies attempting to resolve the injection of particles in relativistic turbulence.  \citet{Comisso2018} show that nonthermal particles in relativistic turbulence feature a sudden increase in energy, representing the injection process (Figure \ref{fig:Comisso}). The figure also shows that this injection is associated with the strong current sheet regions, where electromagnetic fields can change dramatically and reconnection can happen. Using a fitting procedure similar to \citet{French_2023}, \citet{Singh_2024} show that for $\sigma_0 = B_0^2/(4 \pi n_0 m_e c^2) = 20$ and $\delta B \sim B_0$, the injection energy is roughly at $\gamma_{\rm inj} \sim \sigma_0/2$, while the maximum energy scales linearly with the system size.

 \begin{figure}
    \centering
    \includegraphics[width=0.9\textwidth]{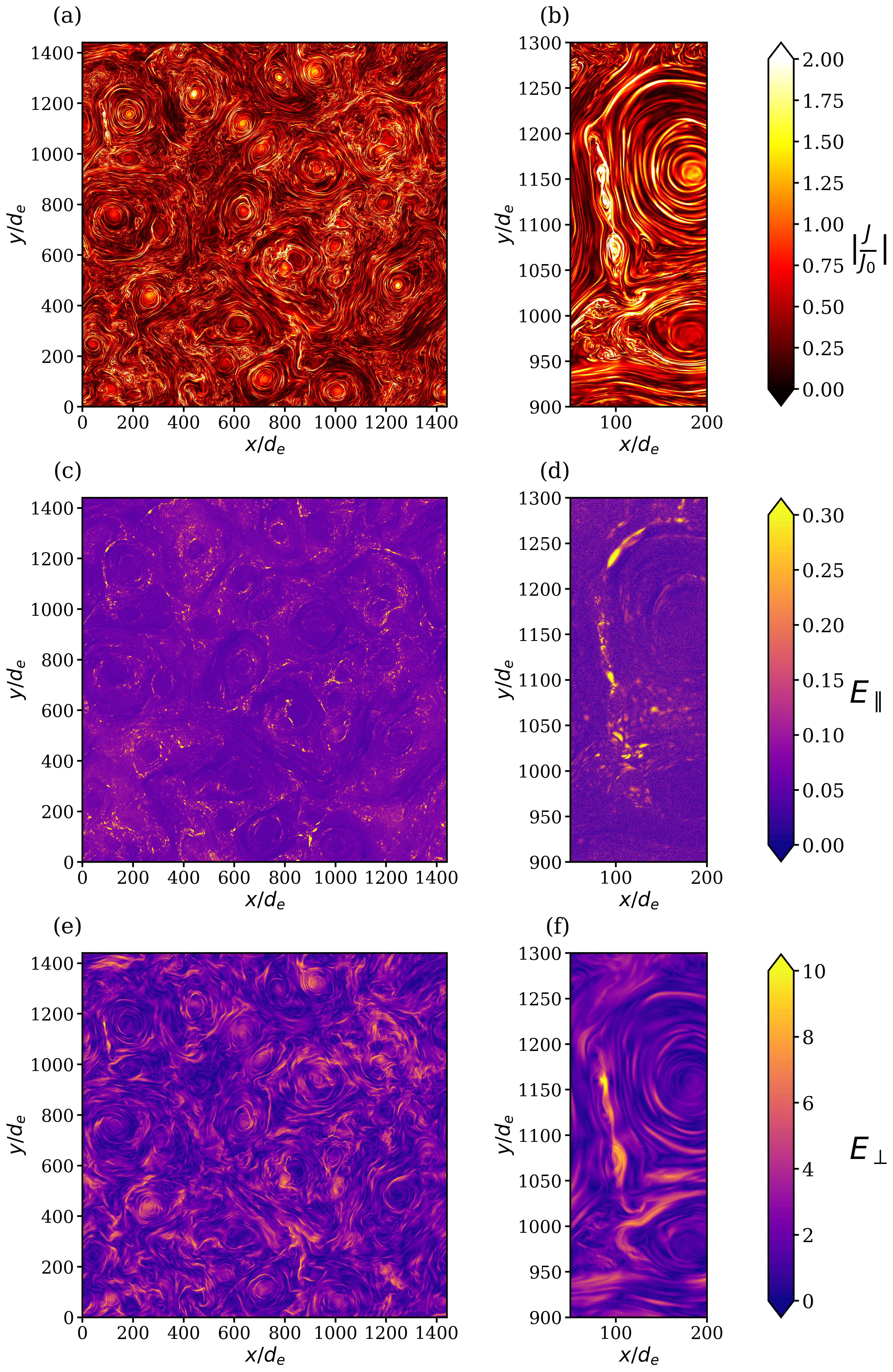}
    \caption{Colormaps of ((a)–(b)) current density magnitude ($|J/J_0|$), ((c)–(d)) parallel electric field ($E_\parallel$), and ((e)–(f)) perpendicular electric fields ($E_\perp$). The right column (panels ((b), (d), and (f))) are zoomed-in versions of the left column (panels ((a), (c), and (e))) that focus on a specific reconnection region around $x/d_e = 100$ and $y/d_e = 1100$. [Adapted from \citet{Singh_2024}, reproduced by permission of the AAS]     \label{fig:Singh_2024_electric_fields} }
\end{figure}

\begin{figure}
    \centering
    \includegraphics[width=\textwidth]{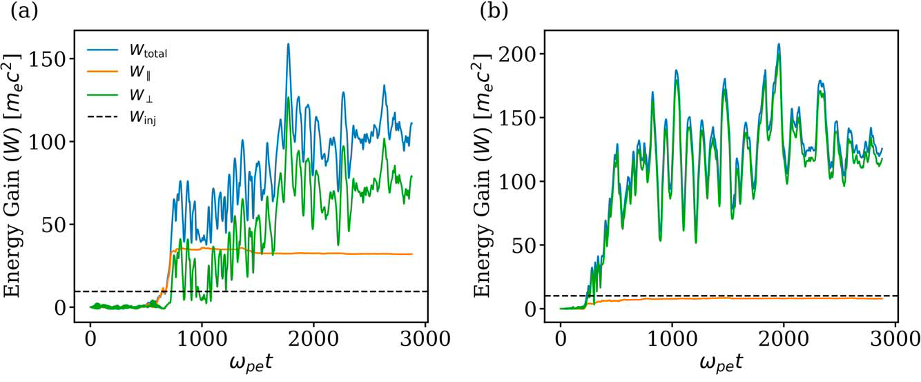}
    \caption{Adapted from~\citet{Singh_2024}. Contributions to total energy gain by $W_\parallel$ and $W_\perp$ for four tracer particles with final energies well above the injection energy. }
    \label{fig:Singh_tracer}
\end{figure}

The mechanism of particle injection has also been investigated in relativistic magnetic turbulence. \citet{Comisso2019} suggested that parallel electric fields in reconnection diffusion regions at the reconnecting current sheets in turbulence dominate the injection process. However, this conclusion mostly focused on highest energy particles in the simulation. As we discussed in earlier sections, the injection process during reconnection itself can involve different mechanisms. Figure \ref{fig:Singh_2024_electric_fields} (adapted from \citet{Singh_2024}) shows the current density and electric fields in a 2D turbulence simulation. Many plasmoids and current sheets are produced in 2D turbulence, where magnetic reconnection is likely to happen. The figure also zooms in on a reconnection site occurring in the simulation. The electric fields are plotted in units of $B_0/\sqrt{2 \sigma_0}$. The $E_\parallel$ is well-localized to reconnection X-points. However, $E_\perp$ does have a substantial strength at reconnection regions owing to the flows near the high current density region. Fig. \ref{fig:Singh_tracer} shows time evolution of tracer particle energy and the contribution of parallel energy gain ($W_\parallel$) and perpendicular energy gain ($W_\perp$). In Figure \ref{fig:Singh_tracer}(a), we see that for a high-energy particle, the
energy gain during injection is dominated by~$W_\parallel$. Later, $W_\parallel$
flattens out, and $W_\perp$ dominates the energy gain. The pattern is
similar to the examples shown in \citet{Comisso2018} and has been seen in reconnection simulations.
The subsequent acceleration for this particle to high
energies is a result of the perpendicular electric fields via
a Fermi-like mechanism. Figure~\ref{fig:Singh_tracer}(b) shows a different
high-energy particle for which $W_\parallel$ flattens out at a much
lower energy and~$W_\perp$ dominates both the injection and
post-injection phases.
 
\begin{figure}
    \centering
    \includegraphics[width=0.6\textwidth]{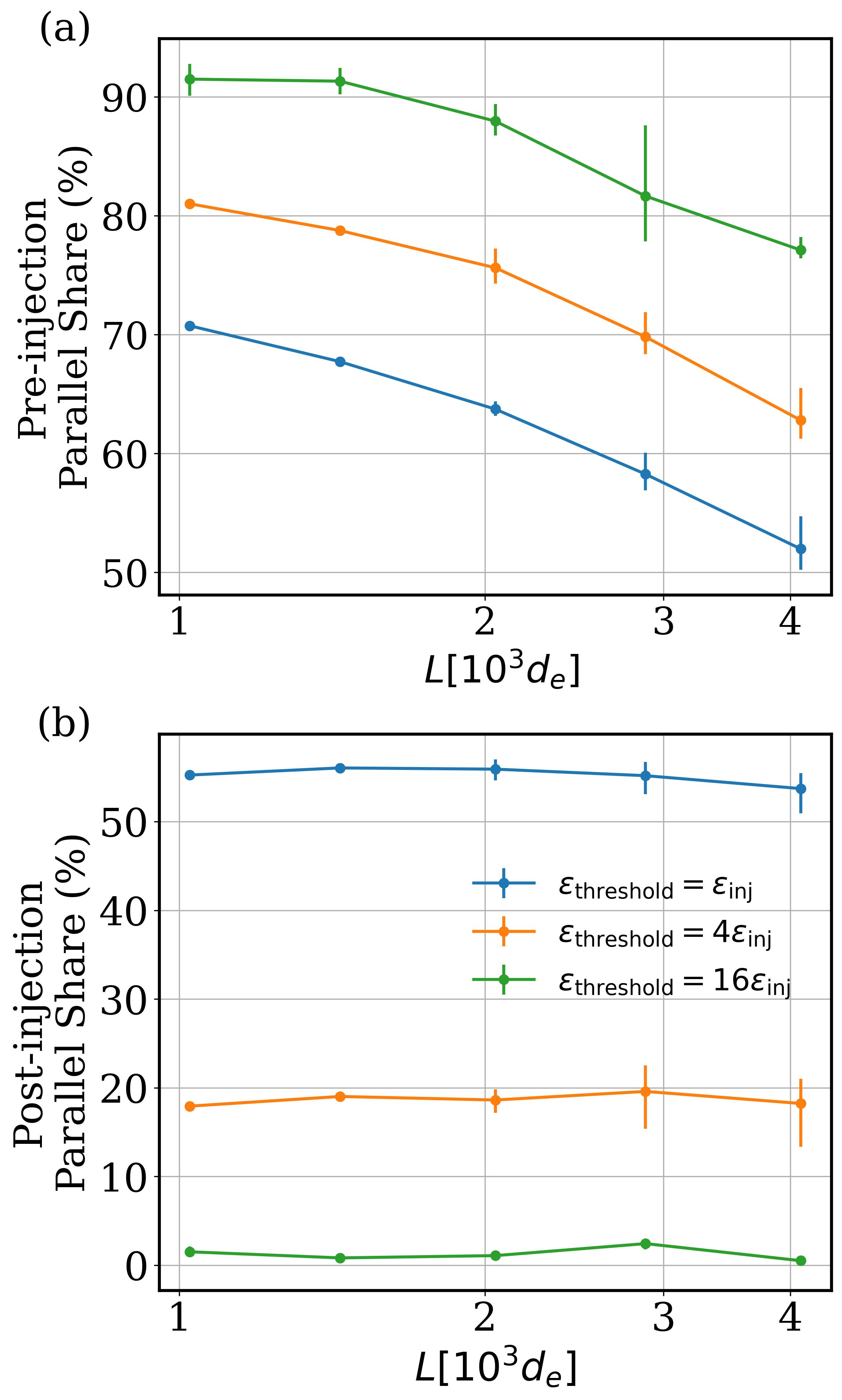}
    \caption{Adapted from~\citet{Singh_2024}, Figure 8. Variation of (a) pre-injection and (b) post-injection share of the work done by the parallel electric field with domain size before and after injection for different injection thresholds.}
    \label{fig:Singh_2024_fig8}
\end{figure}

When extending the analysis to all the particles within the power-law spectrum with $\gamma > \gamma_{inj}$, \citet{Singh_2024} found the roles of~$E_\parallel$ and~$E_\perp$ to be comparable.  In addition, the contribution of the perpendicular electric field increases with the system size, indicating its importance in injecting particles in large-scale systems. Figure \ref{fig:Singh_2024_fig8} shows the contribution of parallel electric field in energy gain during injection and and post-injection acceleration for different energy threshold for counting the injection process. As the system size becomes larger, the contribution from parallel electric field keep decreasing, whereas the post-injection acceleration does not show strong system size dependence. If only focus on the acceleration of highest accelerated particles, parallel electric field dominates the initial acceleration of those particles. However, these particles only consist of a small fraction of particles. As we extend the analysis to all the injected particles, both perpendicular and parallel electric fields are important for injecting particles above $\varepsilon_{inj}$.
\citet{Comisso2019} also show that when a test-particle component do not see the parallel electric field during the simulation, the density of injected particles is dramatically reduced. This indicates that both~$W_\parallel$ and~$W_\perp$ are important for injection in relativistic turbulence with pair plasmas. Because of the background magnetic field, particle injection by small-scale reconnection in turbulence is similar to that in guide field reconnection, with non-ideal electric field more important for $\delta B < B_0$ and motional electric field more important for $\delta B > B_0$. When extending to proton-electron plasmas, injection is more dominated by motional electric field for ions and non-ideal electric field is more important for injecting electrons \cite{Comisso2022}, also consistent with injection process during reconnection \citep{Zhang2024ApJ}.

\section{Discussion and Outlook} \label{sec:discussion}

In this review, we discussed injection problems in various regimes in magnetic reconnection and turbulence. Since different regimes of magnetic reconnection share similar physics, their injection processes share much of the same physics. Also because  magnetic reconnection and turbulence can be intrinsically related, they too share a lot of common physics. Recent studies have found several mechanism can contribute the energization during injection, such as Fermi reflection, direct acceleration, and pickup process.

We would like to emphasize that it is important to be clear about the definition of  particle injection and acceleration. While kinetic simulations from different studies can well be compared with each other, some of the past work only focused on very high end of the energy spectra, but the conclusion can be very different if lower energy particles are included as the nonthermals. By definition, the injection process should mark the beginning of nonthermal acceleration (corresponding to the lower-bound of the power-law energy spectrum). The highest energy end only includes a tiny fraction of the injected particles. To be self-consistent, the primary injection mechanism should be able to explain how the majority of the injected particles are accelerated beyond $\varepsilon_{inj}$.

While the injection problem has been studied by several recent works, the current studies have not filled all the regimes of magnetic reconnection and for all the species. Therefore, a comprehensive understanding is still an ongoing effort. For example, proton acceleration in relativistic reconnection and electron injection during non-relativistic low-beta reconnection are not fully understood.
The role of non-ideal electric field when a guide field is present and when it is proton-electron plasmas. In proton-electron plasma, the electrons nearly shield all the motional electric field and therefore it is difficult for them to gain energy from motional electric field. However, whether the non-ideal electric field can solely accelerate particles to the lower bound of the power-law distribution is still under debate. Meanwhile, it appears that protons and other ion species can be injected via the motional electric field.

As we discussed, the injection stage is important for determining and understanding (i) the relative abundance of and (ii) the energy partition between thermal and nonthermal particles of different species. %Injection process is important for determining the fraction of nonthermal component and understanding the energy partition between thermal and nonthermal energies in different species. It also determines the relative abundance in the energetic particle population.

The insignificant role of the diffusion region with~$E>B$ suggests that one can ignore the kinetic diffusion region for energetic particle models. In fact, models that do not include these but large-scale non-ideal electric field do generate substantial particle injection \citep{Arnold2021}.

While studying the high-energy break of the power-laws require extrapolation, kinetic simulations (fully kinetic and hybrid) are likely able to study the injection process well due to its intrinsic scales. A detailed understanding of the injection mechanisms and their contributions is needed to construct injection models, which are useful in the context of global or large-scale simulations that contain regions that may be well-approximated as extended current sheets \citep{Arnold2021,Li2018b,Seo2024}.

\section*{Acknowledgements} \label{sec:acknowledgements}

We acknowledge the support in part from Los
Alamos National Laboratory through the LDRD program, NASA Award NNH240B72A, 80HQTR21T0087 and NNH24OB107. FG thanks the support from NSF Award 2308091.
OF acknowledges support from the National Science Foundation Graduate Research Fellowship under Grant No. DGE 2040434. QZ acknowledges 80NSSC22K0352 and 80NSSC20K1813.

\section*{Declarations}

\bmhead{Conflict of interest}
The authors have no conflict of interest to declare that is relevant to the content of this article.

%\input{1_intro}
%\input{2_rel_recon}
%\input{3_nonrel_recon}
%\input{4_turb}
%\input{5_discussion}
%\input{6_acknowledgements}

%\bibliographystyle{aipnum4-2} 
%\bibliography{main}

\providecommand{\noopsort}[1]{}\providecommand{\singleletter}[1]{#1}%\providecommand\bibinfo[2]{#2}
%% BioMed_Central_Bib_Style_v1.01

\end{document}